\documentclass[aps,pre,preprint,showpacs,superscriptaddress]{revtex4} 

\usepackage{amssymb}
\usepackage[dvips]{epsfig}
\usepackage{subfigure}
\usepackage{amsmath}

\newcommand{\scl}{0.35}

\begin{document}

\title{Soliton dynamics in symmetric and non-symmetric complex potentials}
\author{Yannis Kominis} 
\address{School of Applied Mathematical and Physical Science,\\
			 National Technical University of Athens,\\
			 Zographou GR-15773, Greece}

\begin{abstract}
Soliton propagation dynamics under the presence of a complex potential are investigated. A large variety of qualitatively different potentials, including periodic, semi-infinite periodic and localized potentials, is considered. Cases of both symmetric and non-symmetric potentials are studied in terms of their effect on soliton dynamics. The rich set of dynamical features of soliton propagation include dynamical trapping, periodic and non-periodic soliton mass variation and non-reciprocal scattering dynamics. These features are systematically investigated with the utilization of an effective particle phase space approach which is shown in remarkable agreement with direct numerical simulations. The generality of the results enables the consideration of potential applications where the inhomogeneity of the gain and loss is appropriately engineered in order to provide desirable soliton dynamics. 

\pacs{42.65.Tg, 42.65.Sf, 03.75.Lm, 05.45.Yv}
\end{abstract}

\maketitle

\section{Introduction}
Soliton formation and dynamics in spatially inhomogeneous structures is a subject of intense research interest with applications to many branches of physics, including optical waves in nonlinear photonic structures \cite{NLO_reviews} and matter waves in Bose-Einstein Condensates (BEC), \cite{BEC_reviews}. Spatial modulations of the linear or the nonlinear refractive index of an optical medium have been shown to result in the formation of self-localized waves that have no counterpart in homogeneous systems. \cite{Lattice_solitons} Lattice solitons have been shown to exist in a large variety of periodic configurations \cite{KoHi_08} whereas surface solitons have been shown to be formed at the interfaces between semi-infinite periodic or inhomogeneous structures \cite{Surface, KoHi_09} and defect solitons are known to be formed at media with localized spatial inhomogeneities. \cite{Defect} In the case of strong spatial modulations, soliton profiles can be interestingly complex but wave dynamics are rather restricted due to the deep soliton trapping and the resulting transverse immobility. Contrarily, in the case of rather weak modulations, soliton profiles remain simple but soliton dynamics can be quite rich and have interesting features with great potential for applications. In such cases solitons move actually as effective particles in a potential, with the form of the latter depending strongly on the characteristics of the soliton. Therefore, different solitons may undergo qualitatively different dynamics in the same inhomogeneous structure. \cite{KoHi_08,KoHi_09} 
        
The consideration of spatial modulation of the material gain and losses appears naturally as a next step for engineering the soliton formation and dynamics and opens new possibilities for applications. The formation of gap solitons has been investigated in periodic lattices with homogeneous \cite{Dis_homogeneous} and inhomogeneous \cite{soliton_PT, Dis_periodic} gain and loss properties. Localized defect \cite{Dis_defect} and surface \cite{Dis_surface} modes have been shown to be supported by localized gain landscapes. \cite{Malomed_MiniReview} In general, soliton stability under the action of such complex potentials is a still open research subject. \

The role of the symmetry of the spatially modulated real and imaginary part of the potential corresponding to the refractive index and material gain/loss has been studied lately in terms of the $\mathcal{PT}$-symmetry which is of interest to quantum mechanics. \cite{Makris_PT, soliton_PT} It has been shown that an even real part and an odd imaginary part of the complex potential is a necessary condition for a purely real spectrum of the respective operator describing wave propagation. In many cases this condition is accompanied with a condition bounding the relative magnitude of the imaginary part with respect to the real part. \cite{Makris_PT} From a practical point of view, a purely real spectrum ensures the stability of the zero background and therefore the immunity of the system to noise, even under large gain modulations. However, in terms of applications, as in most configurations including active media, a relatively small gain results in small growth rates of the background instability and along with a finite length of a device allows for robust wave propagation and negligible noise. Even for cases of homogeneous gain/loss it has been shown that the interplay between the dynamical soliton power variation and the refractive index modulation results in a rich set of soliton dynamical features. \cite{Ko_Dis} \

Cases of asymmetric complex profiles have been quite recently considered  and the existence of continuous families of stationary localized nonlinear modes has been shown, \cite{asymmetric_1} whereas symmetry-breaking of solitons has been shown to occur even in $\mathcal{PT}$-symmetric potentials. \cite{asymmetric_2} The existence of continuous families of solitons has been shown to be related to "hidden" symmetries of soliton formation dynamics that are not necessarily related to the $\mathcal{PT}$-symmetry of the potential. These results suggest that a wide class of asymmetric complex potentials can support stationary solitons and significantly extend the range of possible applications.\     

In this work we study soliton dynamics under the presence of relatively weak symmetric and non-symmetric complex potentials for a large variety of spatial modulations including periodic and localized inhomogeneities as well as interfaces between homogeneous and semi-infinite periodic media. In contrast to most of the previous studies, the emphasis is not the formation of stationary solitons but on soliton dynamics under propagation. The presence of gain and loss not only affects the soliton mass (power) but also the effective potential under which the soliton is moving due to the spatial modulation of the refractive index. Soliton dynamics are studied in the three-dimensional phase space of an effective particle of varying mass and the role of spatial symmetries as well as deviations from symmetry is investigated.

\section{Model} 
Soliton propagation in the presence of a complex potential is described by the inhomogeneous NLS equation:
\begin{equation}
iu_z+u_{xx}+\left[V(x)+iW(x)\right]u+2|u|^2u=0 \label{NLS}
\end{equation}
where $u$ is the wave field envelope, $z$ the normalized propagation distance, and $x$ the scaled transverse coordinate. $V(x)$ and $W(x)$ are the real and imaginary parts of the complex potential. The soliton can be treated as an effective particle \cite{effective_particle} of variable mass $m=\int|u|^2dx$ and momentum $p=i\int(uu_x^*-u_xu^*)dx=mv$ at a position $x_0$, corresponding to soliton's center, moving with velocity $v$ in an effective potential $U_{eff}$ due to the actual complex potential, according to the equations  
\begin{eqnarray}
\frac{dm}{dz}&=&2\int_{-\infty}^{+\infty}|u|^2W(x)dx \label{dmdz}\\
m\frac{dv}{dz}&=&-\frac{\partial}{\partial x_0}\left[2\int_{-\infty}^{+\infty}|u|^2 V(x)dx \right] \equiv -\frac{\partial U_{eff}}{\partial x_0} \label{dvdz} \\
\frac{dx_0}{dz}&=&v \label{dx0dz}
\end{eqnarray}
The dynamical system defined by Eqs. (\ref{dmdz})-(\ref{dx0dz}) determines soliton dynamics. In the case of real potential ($W=0$), soliton moves with a constant mass, the system has fixed points at the extrema of the effective potential ($U_{eff}$) and the total energy of the effective particle $H=mv^2/2+U_{eff}(x_0)$ is conserved. The presence of a nonzero part of the potential ($W\neq0$) introduces an additional degree of freedom related to the particle mass variation and causes the destruction of the conserved quantity of total energy. These features result in drastic qualitative changes of soliton dynamics in comparison to cases of real potentials.\

Fixed point of the system (\ref{dmdz})-(\ref{dx0dz}) are given by
\begin{eqnarray}
\int_{-\infty}^{+\infty}|u(x-x_0)|^2W(x)dx &=&0  \label{W} \\  
\frac{\partial}{\partial x_0}\int_{-\infty}^{+\infty}|u(x-x_0)|^2V(x)dx &=&0 \\
v&=&0
\end{eqnarray}
and correspond to solitons propagating with a fixed mass at a fixed position (zero velocity). From these equations we conclude that fixed points occur at positions $x_0$ with respect to which $W(x)$ is odd and $V(x)$ is even, similarly to the necessary condition for $\mathcal{PT}$-symmetry. \cite{Makris_PT} Note that the condition (\ref{W}), corresponding to the requirement of balance between gain and loss, does not introduce a constraint on the soliton mass (and the respective propagation constant), contrary to the general case of dissipative systems. \cite{AA_Dis}   \

Moreover, from Eqs. (\ref{dmdz}),(\ref{dvdz}) we obtain
\begin{equation}
m\frac{dv}{dm}=\frac{-\frac{\partial}{\partial x_0}\left[\int_{-\infty}^{+\infty}|u|^2 V(x)dx \right]}{\int_{-\infty}^{+\infty}|u|^2W(x)dx}\equiv I(m,v,x_0)
\end{equation}      
with the quantity $I(m,v,x_0)$ depending, in general, on all soliton parameters. However, it is readily seen that under the condition
\begin{equation}
\frac{\partial V(x)}{\partial x}=C W(x) \label{int_condition}
\end{equation}
with $C$ being a constant, we have $I(m,v,x_0)=-C$ resulting in a conserved quantity of the effective particle motion given by
\begin{equation}
K(m,v)=C\ln m + v =\mbox{const.} \label{K}
\end{equation}
that restricts soliton dynamics in two-dimensional submanifolds of the phase space $(x_0,v,m)$. Moreover, this condition implies the existence of a stable/unstable fixed point at the minima/maxima of the real part of the potential. \

It is worth emphasizing the different restrictions imposed on the real and imaginary parts of the potential by the different conditions for the existence of a fixed point and a constant of the motion. \emph{A fixed point at $x_0$ exists whenever the real and the imaginary part are even and odd, respectively, independently of any functional relation between them, whereas the condition (\ref{int_condition}) imposes a mutual restriction in the form of a functional relation between the real and the imaginary part of the potential but does not impose any restriction on their spatial symmetry.} Note that, the condition for the existence of the conserved quantity is qualitative different from the necessary condition for $\mathcal{PT}$-symmetry. However, under the condition (\ref{int_condition}), when $V(x)$ is even, $W(x)$ is odd and vice versa. \   

The above conditions and discussion are generic with respect to the amplitude and the profile of the complex potential, since the equations (\ref{dmdz}) and (\ref{dvdz}) are exact equations for the soliton mass and velocity variation under propagation when $u$ is an exact solution of eq. (\ref{NLS}). However, in this work we are mostly interested in soliton dynamics that occur in relatively weak potentials where the solitons are quite mobile. In this case the equations (\ref{dmdz}) and (\ref{dvdz}) can be treated perturbatively and provide analytical results by utilizing in the respective integrals the well known soliton solution of the homogeneous NLS equation ($V=W\equiv0$) that is given by
\begin{equation}
u=\eta\mbox{sech}[\eta(x-x_0)]\exp[i(vx/2+2\phi)] \label{NLS_sol}
\end{equation}
with $x_0$ and $v=dx_0/dz$ being the position and the velocity of the soliton center and $d\phi/dz=\eta^2/2-v^2/8$. The soliton mass is $m=2\eta$.\ 

In the following, we focus on three characteristic categories of potential profiles, that is a periodic profile, a localized defect and a semi-infinite periodic profile interfaced with a homogeneous part. We investigate soliton dynamics for cases where symmetry conditions or the the condition (\ref{int_condition}) are either fulfilled or violated. In all cases, the amplitude of the various potentials are of the order of $10^{-2}$ so that the perturbative approach is valid. The analytical results are compared with numerical simulations of the NLS equation (\ref{NLS}) where a random noise of magnitude $1\%$ with respect to the soliton amplitude has been superimposed to the initial conditions in order to take into account the zero background instability.

\section{Periodic potential} 
A characteristic periodic profile of the complex potential is the sinusoidal profile 
\begin{eqnarray}
V(x)&=&V_0\cos\left(K_0x+\Delta x\right) \nonumber \\
W(x)&=&W_0\sin\left(L_0x\right) \label{periodic}
\end{eqnarray}
with $V_0$, $W_0$ being the amplitudes and $K_0$, $L_0$ the wavenumbers of the real and imaginary parts of the potential. The real part of the potential is an even function for $\Delta x=0$. The complex potential is known to have a purely real spectrum under the additional condition $W_0<V_0$ for $K_0=L_0$. \cite{Makris_PT} The condition (\ref{int_condition}) for the existence of the invariant quantity (\ref{K}) requires both $\Delta x=0$ and $K_0=L_0$ but does not restrict the relative amplitude of the real and imaginary parts. For the potential (\ref{periodic}), Eqs. (\ref{dmdz}), (\ref{dvdz}) provide
\begin{eqnarray}
\frac{dm}{dz}&=&-\frac{2\pi W_0 L_0}{\mbox{sinh}\left(L_0\pi/m\right)} \sin(L_0x)\\
m\frac{dv}{dz}&=&-\frac{\partial U_{eff}}{\partial x_0} 
\end{eqnarray} 
with
\begin{equation}
U_{eff}=-\frac{2\pi V_0 K_0}{\mbox{sinh}\left(K_0\pi/m\right)} \cos\left(K_0 x+\Delta x\right)
\end{equation}
Soliton moves as a particle of varying mass in a potential having a constant spatial period but dynamically varying amplitude due to its strong dependence on the particle mass. The topology of the orbits in the three-dimensional phase space $(x_0,v,m)$ depend strongly on the parameters of the potential as shown in Fig. 1. The case of an even real part and an odd imaginary part with equal periods $(L_0=K_0)$ is shown in Fig. 1(a) for soliton initial conditions corresponding to $m=1$, positive and negative velocities $(v)$ and various positions $(x_0)$. It is obvious that, in contrast to the conservative case $W_0=0$, initial conditions with $x_0$ and $v$ of opposite sign do not follow the same orbit. Moreover, all orbits with the same initial mass and velocity are restricted on the two-dimensional invariant manifold (\ref{K}), due to the fulfillment of the condition (\ref{int_condition}) as shown in Fig. 1(b). Characteristic cases of trapped and traveling soliton propagation are shown in Fig. 2(a) and (b). It is worth emphasizing that in the case of a conservative potential the soliton amplitude and width oscillate in such a way that the soliton mass remain constant, whereas in the dissipative case the soliton mass undergoes oscillations.\

Phase space orbits for the case of a potential with an even real and an odd imaginary part, but with different periods of a rational ratio, are depicted in Figs. 1(c) and (d) for positive and negative initial velocities, respectively. In this case the condition (\ref{int_condition}) is not fulfilled and orbits are not restricted in a two-dimensional manifold. Moreover, as shown in Fig. 1(d), the soliton mass variation can be nonperiodic. Soliton propagation for such a characteristic case of continuous mass increasing is depicted in Fig. 2(c).
The case of real and imaginary parts with spatial periods of an irrational ratio is depicted in Fig. 1(e). It is shown that in addition to trapped orbits, we also have orbits corresponding to traveling solitons with quasiperiodic mass oscillations, each one densely filling a two-dimensional surface. Soliton propagation for such a characteristic case is shown in Fig. 2(d), where the inset shows the details of the quasiperiodic mass and amplitude oscillations.\

Finally, a case where neither a spatial symmetry exist nor the condition (\ref{int_condition}) is fulfilled is shown in Fig. 1(e), where the real part is not an even function whereas the imaginary part is an odd function. In this case, there exist an initial condition for which  $x_0=-\Delta x$ and $v=0$ remain constant but the local loss is nonzero, resulting to a soliton evolution where the soliton mass continuously decreases and no transverse soliton motion takes place. Such a characteristic case is depicted in Fig. 2(e). Other initial conditions can result to traveling solitons with increasing mass or trapped solitons with decreasing mass, as also shown in Fig. 2(f).    

\begin{figure}[h]	
\begin{center}
	\subfigure[]{\scalebox{\scl}{\includegraphics{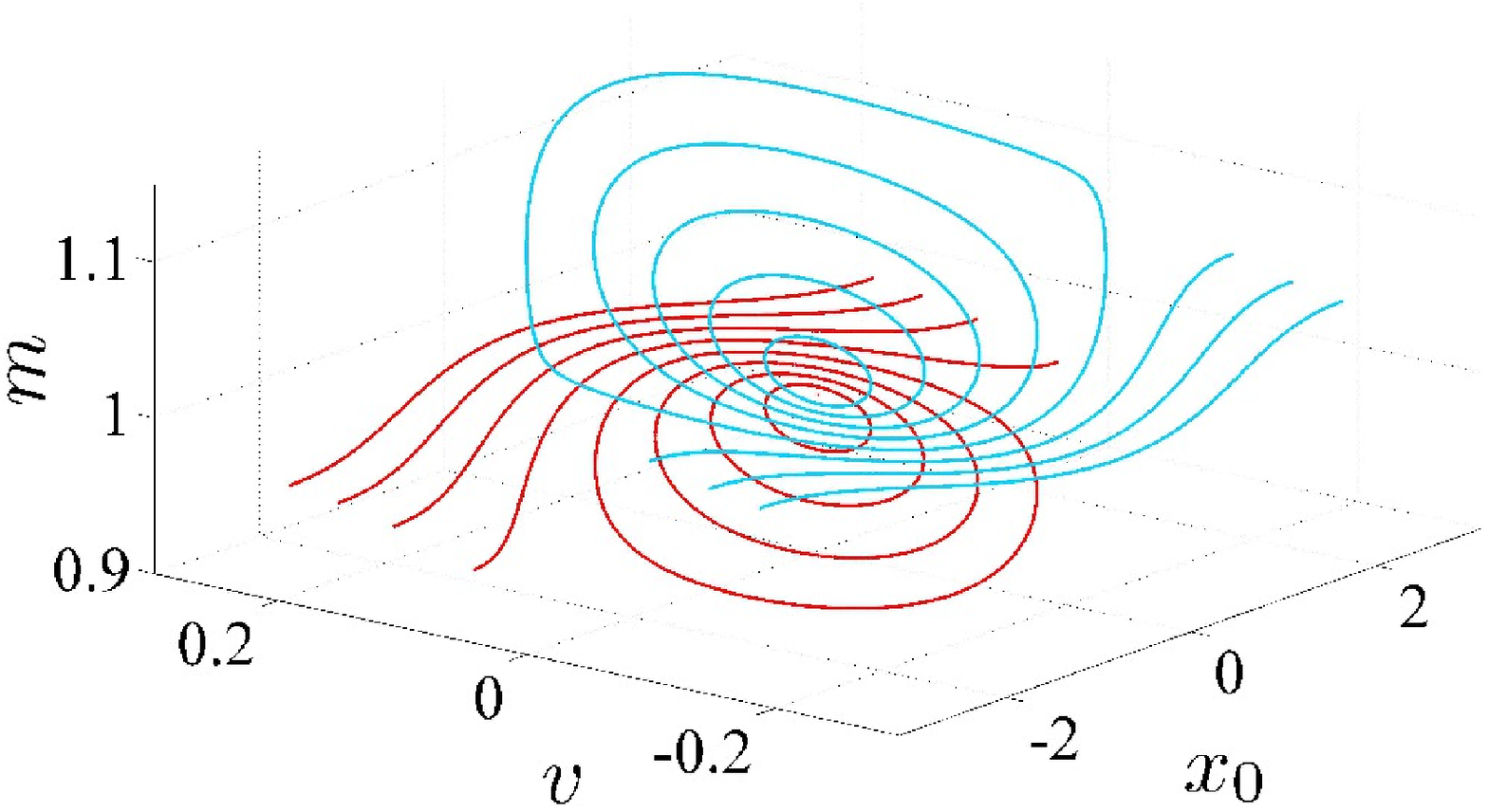}}}
	\subfigure[]{\scalebox{\scl}{\includegraphics{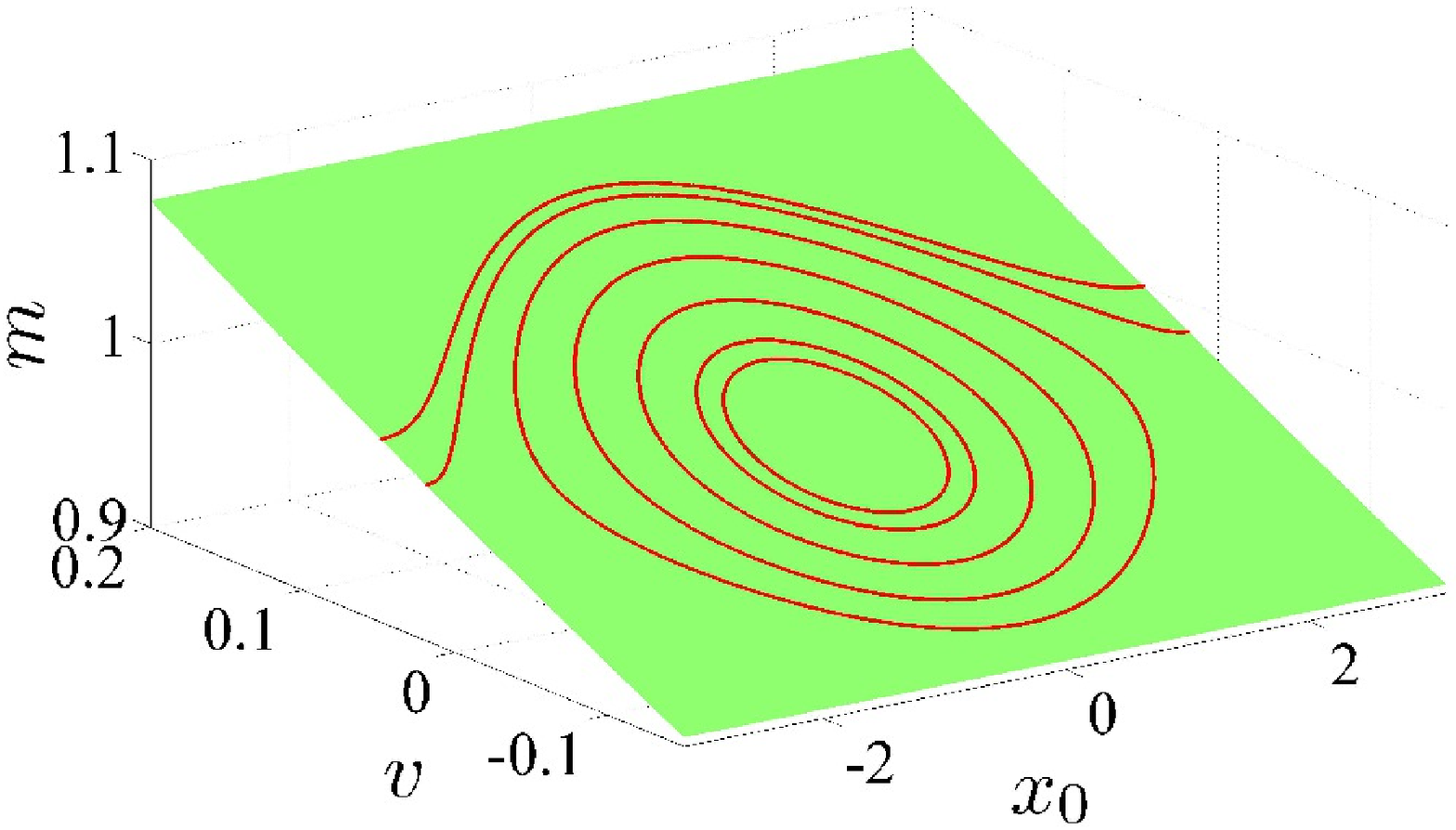}}}
 	\subfigure[]{\scalebox{\scl}{\includegraphics{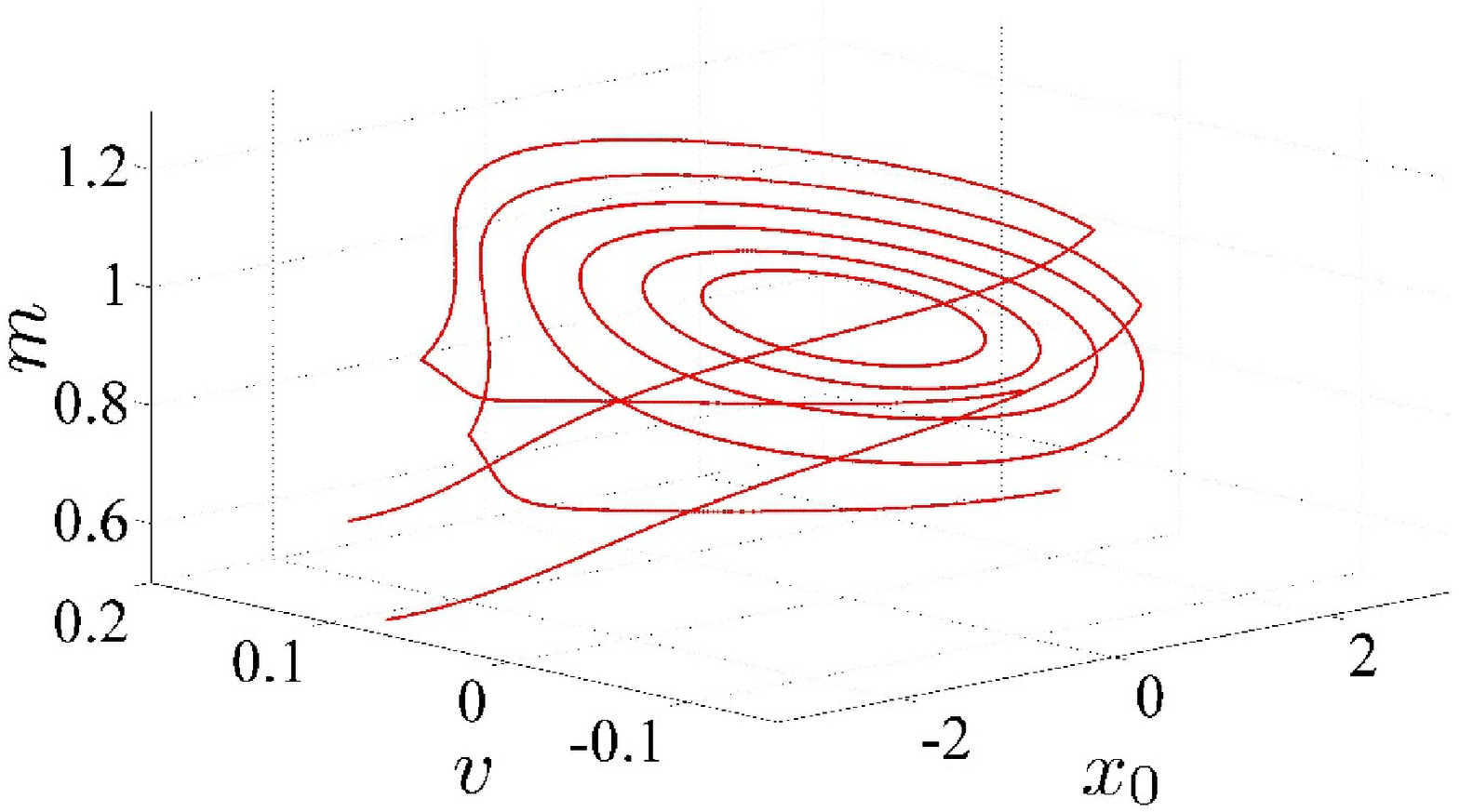}}}
 	\subfigure[]{\scalebox{\scl}{\includegraphics{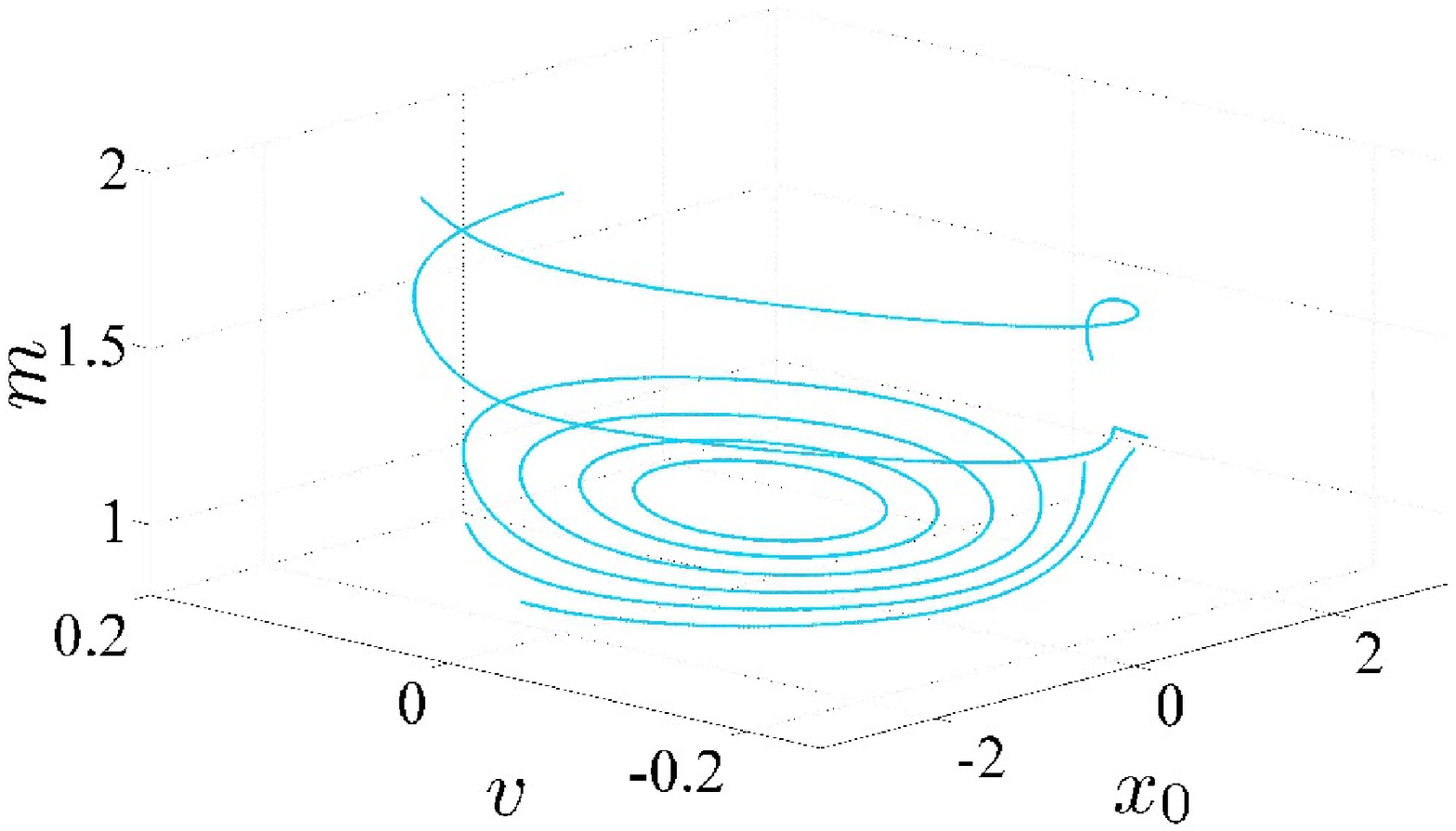}}}
	\subfigure[]{\scalebox{\scl}{\includegraphics{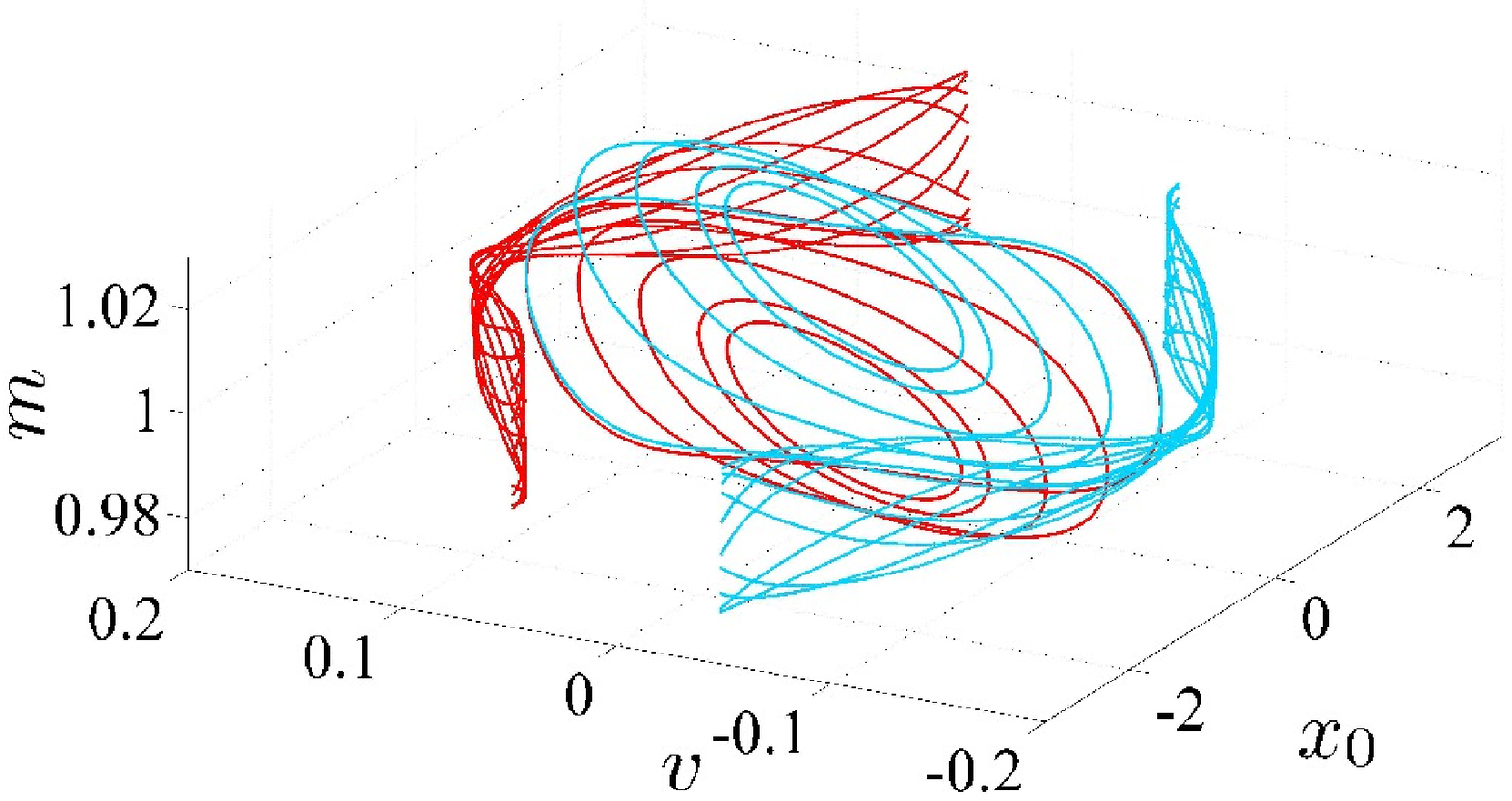}}} 	
	\subfigure[]{\scalebox{\scl}{\includegraphics{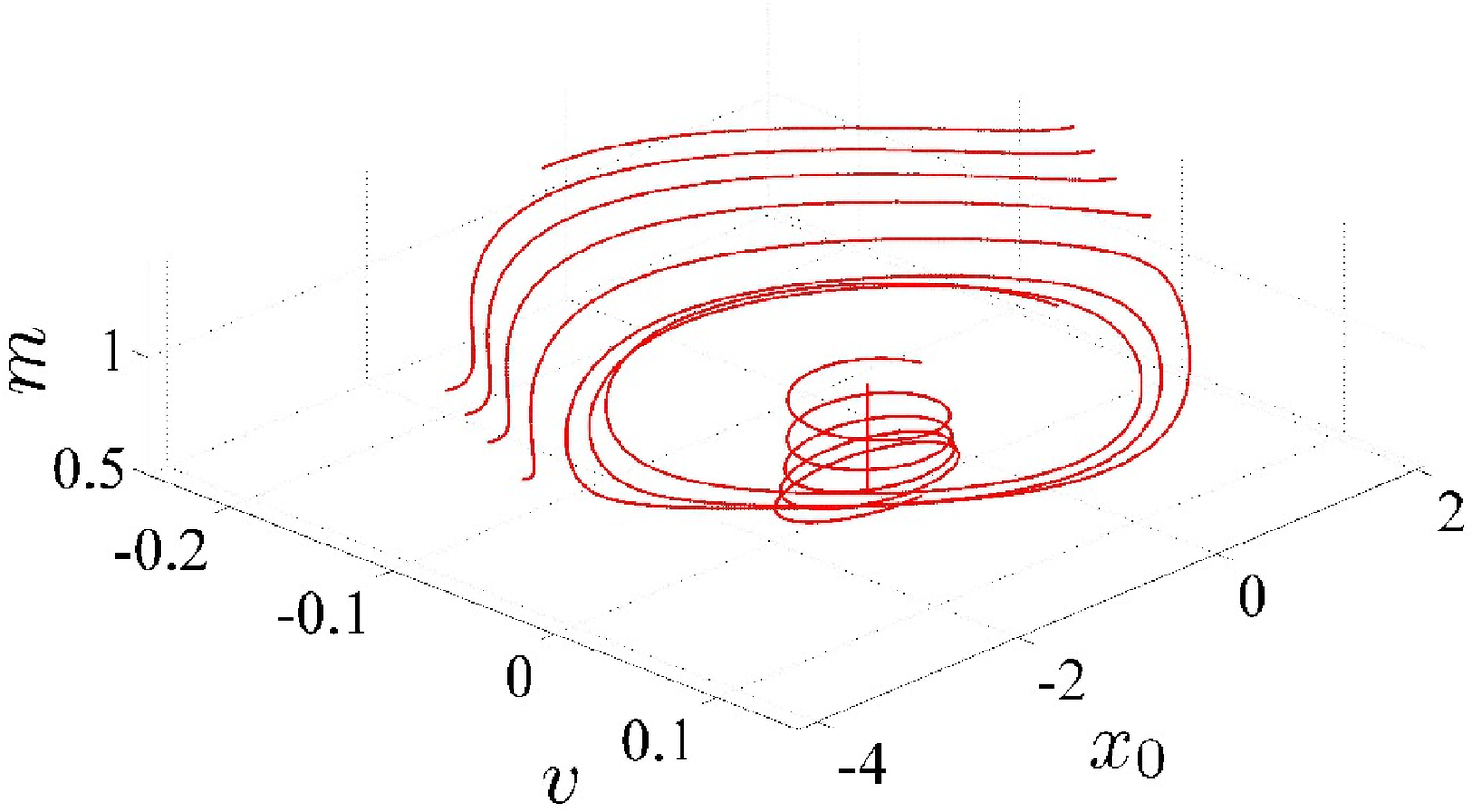}}}
	\caption{(Color online) Phase space orbits of the effective particle model for a soliton with initial mass $m(0)=1$ in the periodic potential (\ref{periodic}) with $V_0=0.01$, $W_0=V_0/2$ and $K_0=1$. (a) Potential: $L_0=1$, $\Delta x=0$, Initial conditions: $x_0(0)=0$, $v(0)>0$ (red / dark gray), $v(0)<0$, (cyan / light gray); (b) Potential: $L_0=1$, $\Delta x=0$, Initial conditions: $v(0)=0.05$ (the two-dimensional surface (\ref{K}) is also shown); (c) Potential: $L_0=1/3$, $\Delta x=0$, Initial conditions: $v(0)=0.05$; (d) Potential: $L_0=1/3$, $\Delta x=0$, Initial conditions: $v(0)=-0.05$; (e) Potential: $L_0=\sqrt{2}$, $\Delta x=0$, Initial conditions: $v(0)>0$ (red / dark gray), $v(0)<0$, (cyan / light gray); (f) Potential: $L_0=1$, $\Delta x=-\pi/3$, Initial conditions: $v(0)=0$, $x_0(0)=\pi/3, 1.5\pi/3, 2.8\pi/3$.}
	\label{Fig:1}
\end{center}
\end{figure}

\begin{figure}[h]	
\begin{center}
	\subfigure[]{\scalebox{\scl}{\includegraphics{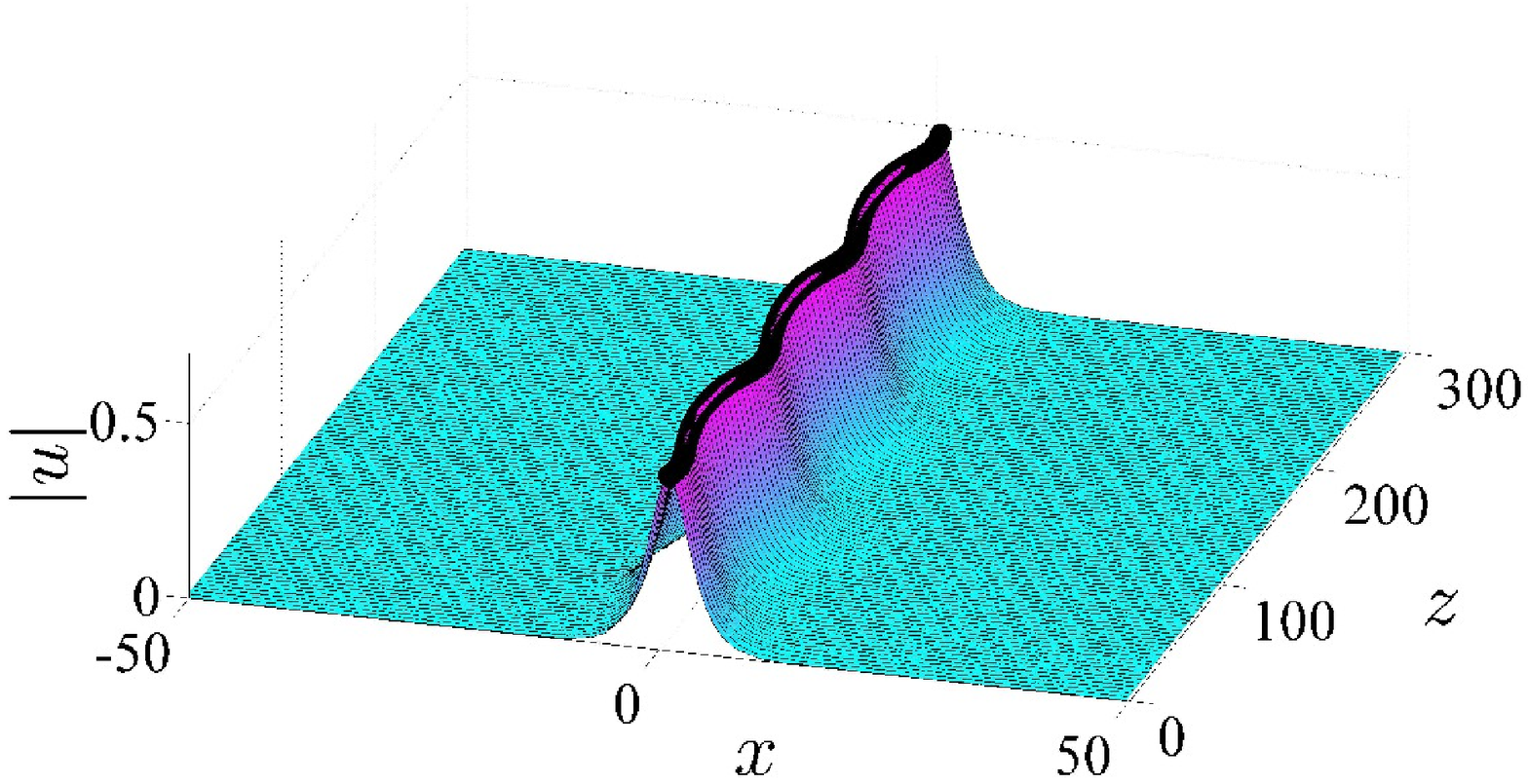}}}
 	\subfigure[]{\scalebox{\scl}{\includegraphics{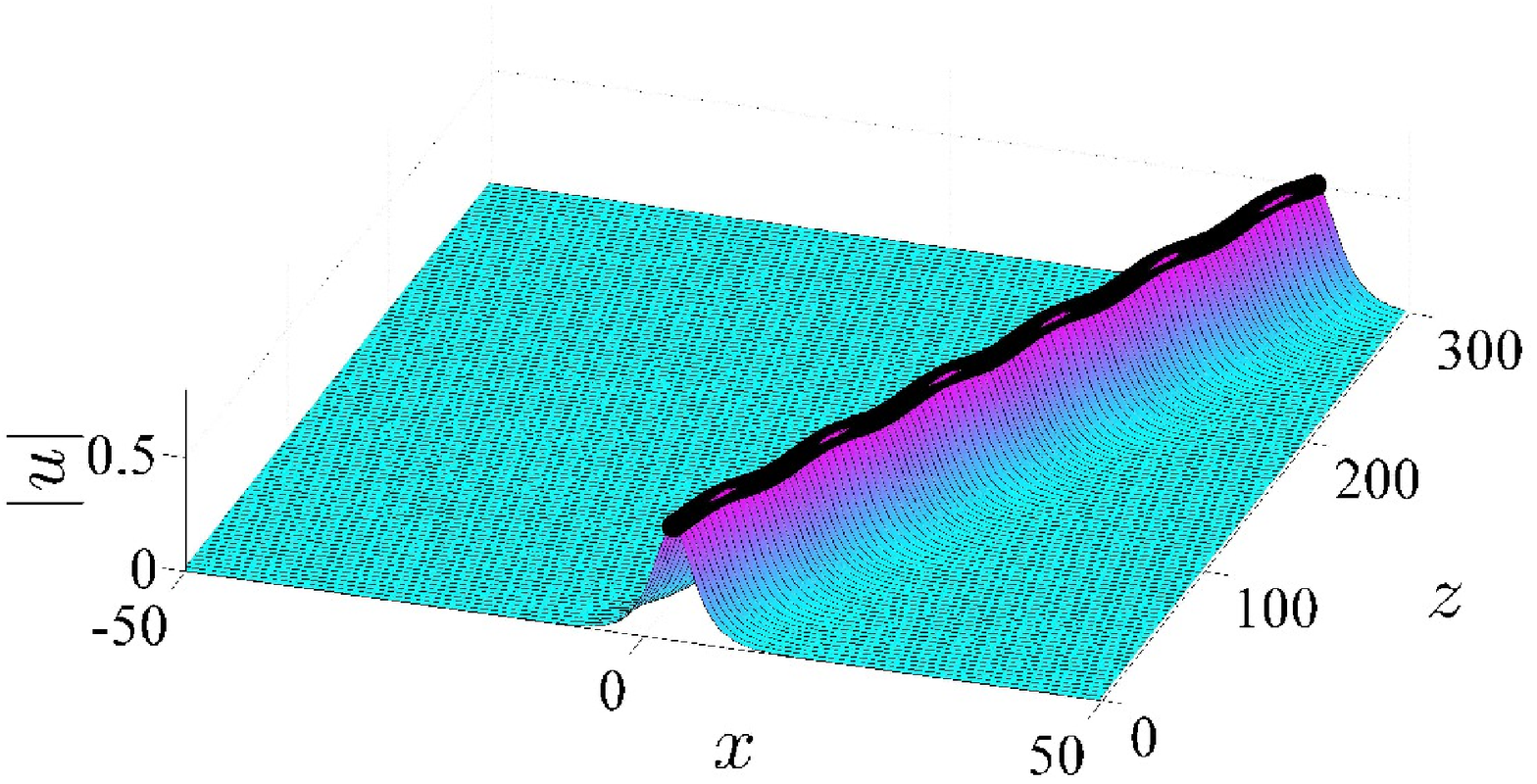}}}
 	\subfigure[]{\scalebox{\scl}{\includegraphics{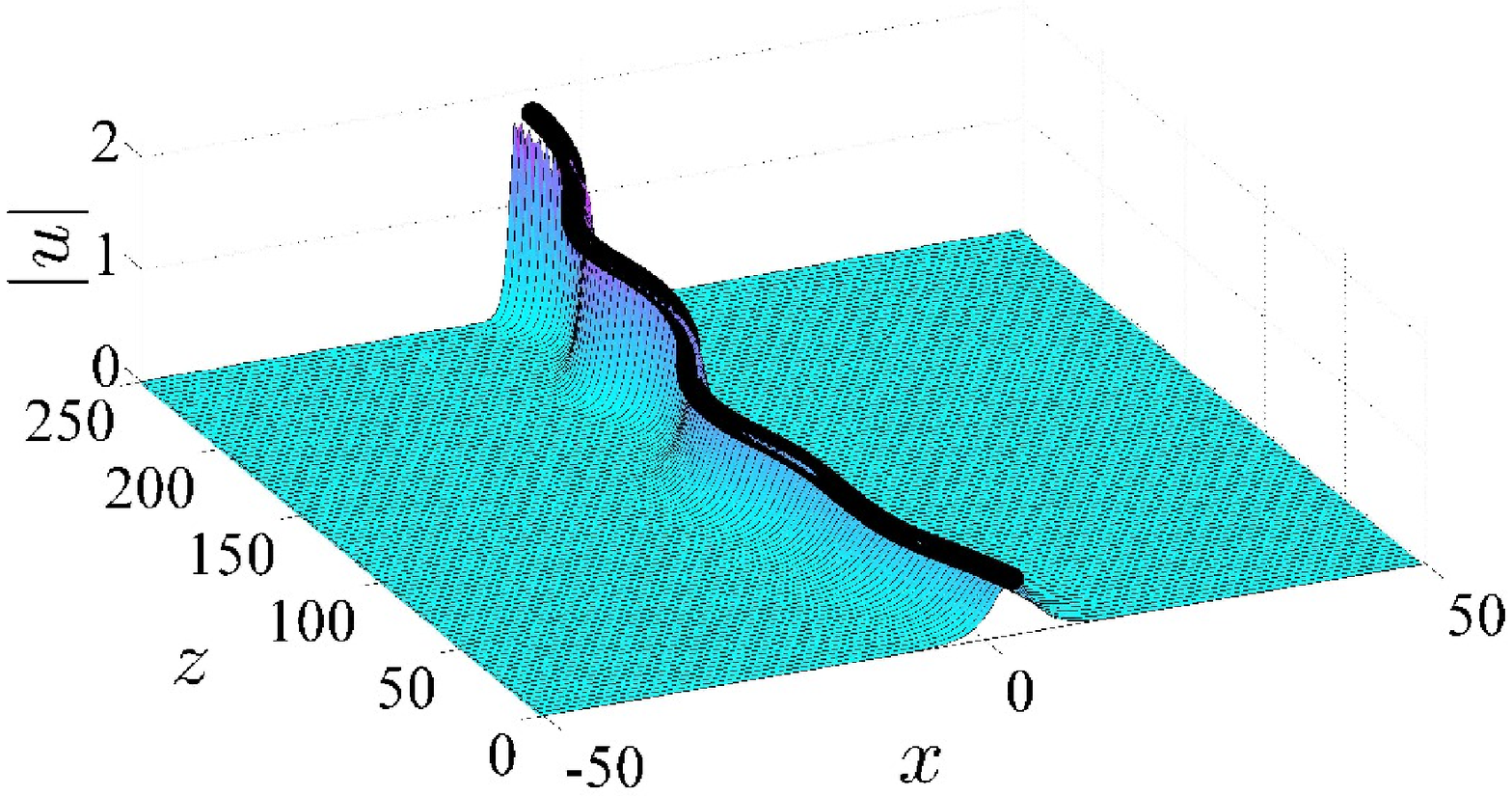}}}
 	\subfigure[]{\scalebox{\scl}{\includegraphics{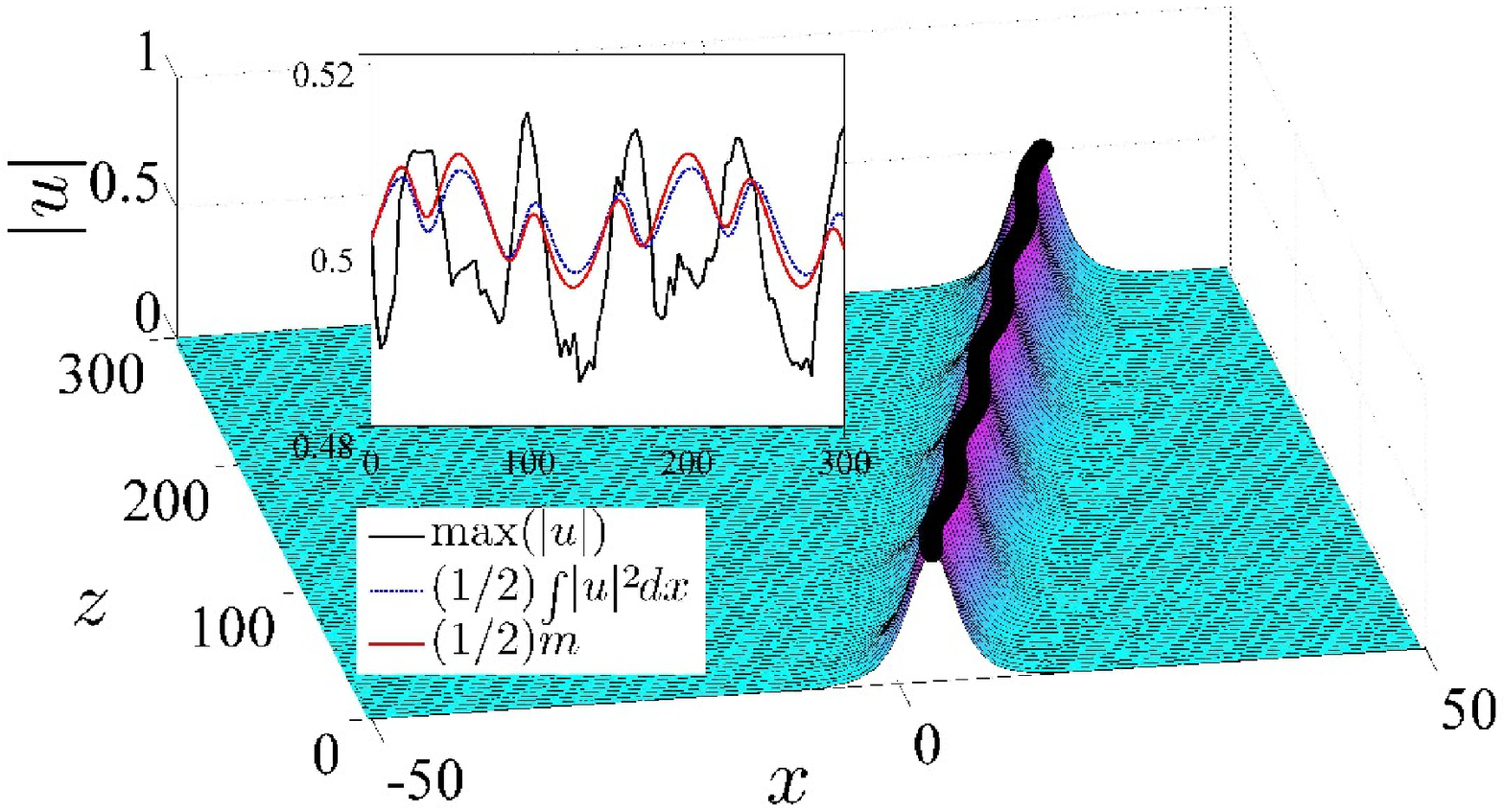}}}
 	\subfigure[]{\scalebox{\scl}{\includegraphics{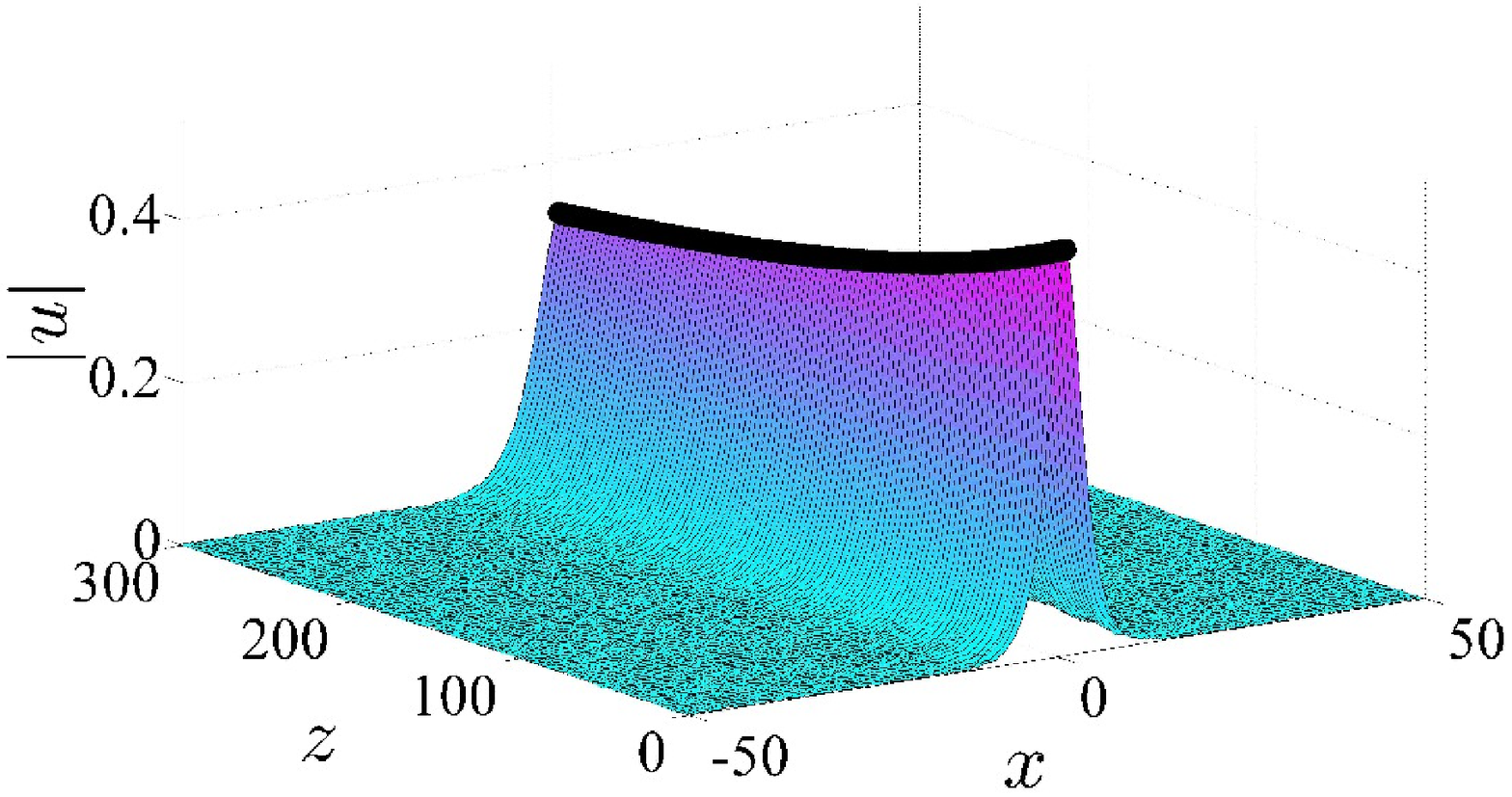}}}
 	\subfigure[]{\scalebox{\scl}{\includegraphics{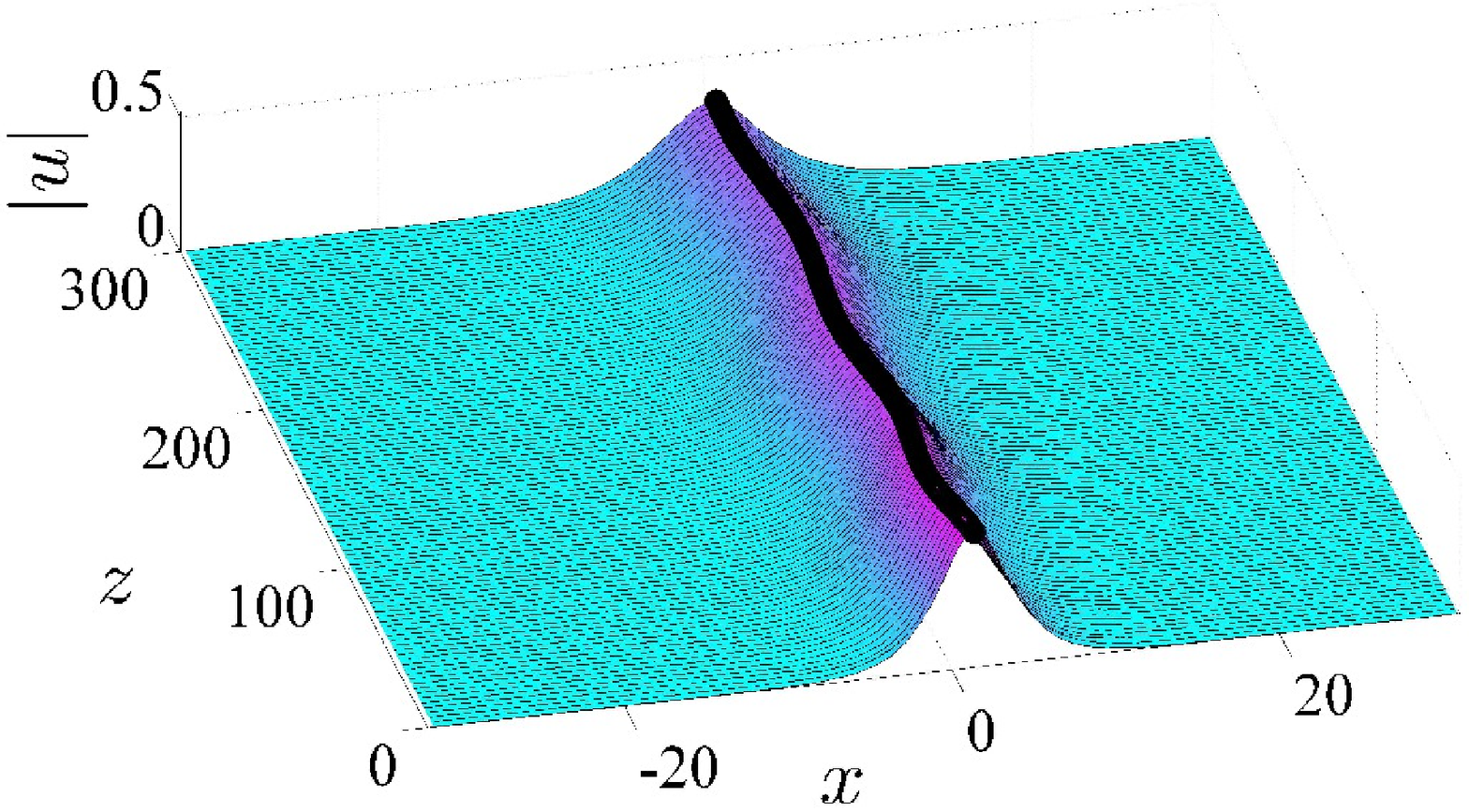}}}
 	\caption{(Color online) Soliton evolution under propagation as obtained from numerical simulations of the NLS equation (\ref{NLS}) and the effective particle model. The thick black line depicts the effective particle orbit $(x_0, v, m/2)$. (a) Potential: as in Fig. 1(a),(b), Initial conditions: $m(0)=1$, $x_0(0)=1$, $v(0)=0.02$;  (b) Potential: as in Fig. 1(a),(b), Initial conditions: $m(0)=1$, $x_0(0)=\pi$, $v(0)=0.08$; (c) Potential: as in Fig. 1(c),(d), Initial conditions: $m(0)=1$, $x_0(0)=\pi$, $v(0)=-0.05$;  (d) Potential: as in Fig. 1(e), Initial conditions: $m(0)=1$, $x_0(0)=\pi$, $v(0)=0.05$, The inset depicts $\max_x(|u|)$ (black), $(1/2)\int|u|^2dx$ (red) and $m/2$ (blue); (e) Potential: as in Fig. 1(f), Initial conditions: $m(0)=1$, $x_0(0)=\pi/3$, $v(0)=0$; (e) Potential: as in Fig. 1(f), Initial conditions: $m(0)=1$, $x_0(0)=1.3\pi/3$, $v(0)=0$.}
	\label{Fig:2}
\end{center}
\end{figure}

\section{Localized potential}
A characteristic case of a complex localized potential is one corresponding to a defect such as
\begin{eqnarray}
V(x)&=&V_0\mbox{sech}\left[K_0(x+\Delta x)\right] \nonumber \\
W(x)&=&W_0\mbox{sech}\left(L_0x\right)\mbox{tanh}\left(L_0x\right) \label{localized}
\end{eqnarray}
where $V_0$, $W_0$ are the amplitudes, and $K_0$, $L_0$ determine the spatial widths, of the real and imaginary parts of the defect. Analogously to the periodic case, the imaginary part is an odd function and the real part is an even function when $\Delta x=0$ whereas the stronger condition (\ref{int_condition}) for the existence of the invariant quantity (\ref{K}) additionally requires $K_0=L_0$. It is worth emphasizing that the quite similar Scarff II potential, widely used in studies of soliton dynamics at complex localized defects, \cite{Makris_PT, soliton_PT} does not fulfill the condition (\ref{int_condition}). For the case of the complex potential (\ref{localized}) an explicit analytical form of the equations of the motions for the effective particle cannot be provided. However, the respective integrals can be calculated numerically and the equations (\ref{dmdz})-(\ref{dx0dz}) with $u$ given by Eq. (\ref{NLS_sol}) can be used. The respective effective potential is localized around $x_0=\Delta x$ and forms either a well $(V_0>0)$ or a barrier $(V_0<0)$ and its amplitude depends on the relation between the soliton mass and the spatial width of $V(x)$ analogously to the periodic case. In the following we focus in the case of a barrier and investigate soliton scattering dynamics by such a defect.\

Phase space orbits for the case where the real part is an even function and the imaginary part is odd function with $L_0=K_0$ is depicted in Fig. 3(a), where reflected and transmitted soliton orbits are shown, corresponding to initial velocities lower and higher than the velocity threshold required to overcome the barrier. The initial conditions correspond to solitons initially located at the left/right of the defect with positive/negative velocities. Due to the presence the imaginary part of the potential the scattering process is not reciprocal so that soliton reflection or transmission depends on the direction. Moreover, soliton mass variation takes place during interaction with the complex defect, so that the transmitted or reflected soliton mass is different from the initial mass. For the case of initial conditions of the same mass and velocity and different position, the respective orbits are restricted in the two-dimensional invariant manifold (\ref{K}) as shown in Fig. 3(b). Characteristic cases of soliton reflection and transmission are depicted in Figs. 4(a) and (b).\

The case where the real part is even and the imaginary part is odd with $L_0 \neq K_0$ is depicted in Fig. 3(c). In this case the condition (\ref{int_condition}) is not fulfilled and the respective orbits are not restricted in a two-dimensional surface, allowing for larger mass variations. The non-reciprocal character of soliton reflection in such case is depicted in Figs. 4(c) and (d). Finally, phase space orbits, for the case $\Delta x \neq 0$, where neither the real part is an even function nor the condition (\ref{int_condition}) is fulfilled, are shown in Fig. 3(d). 

\begin{figure}[h]	
\begin{center}
	\subfigure[]{\scalebox{\scl}{\includegraphics{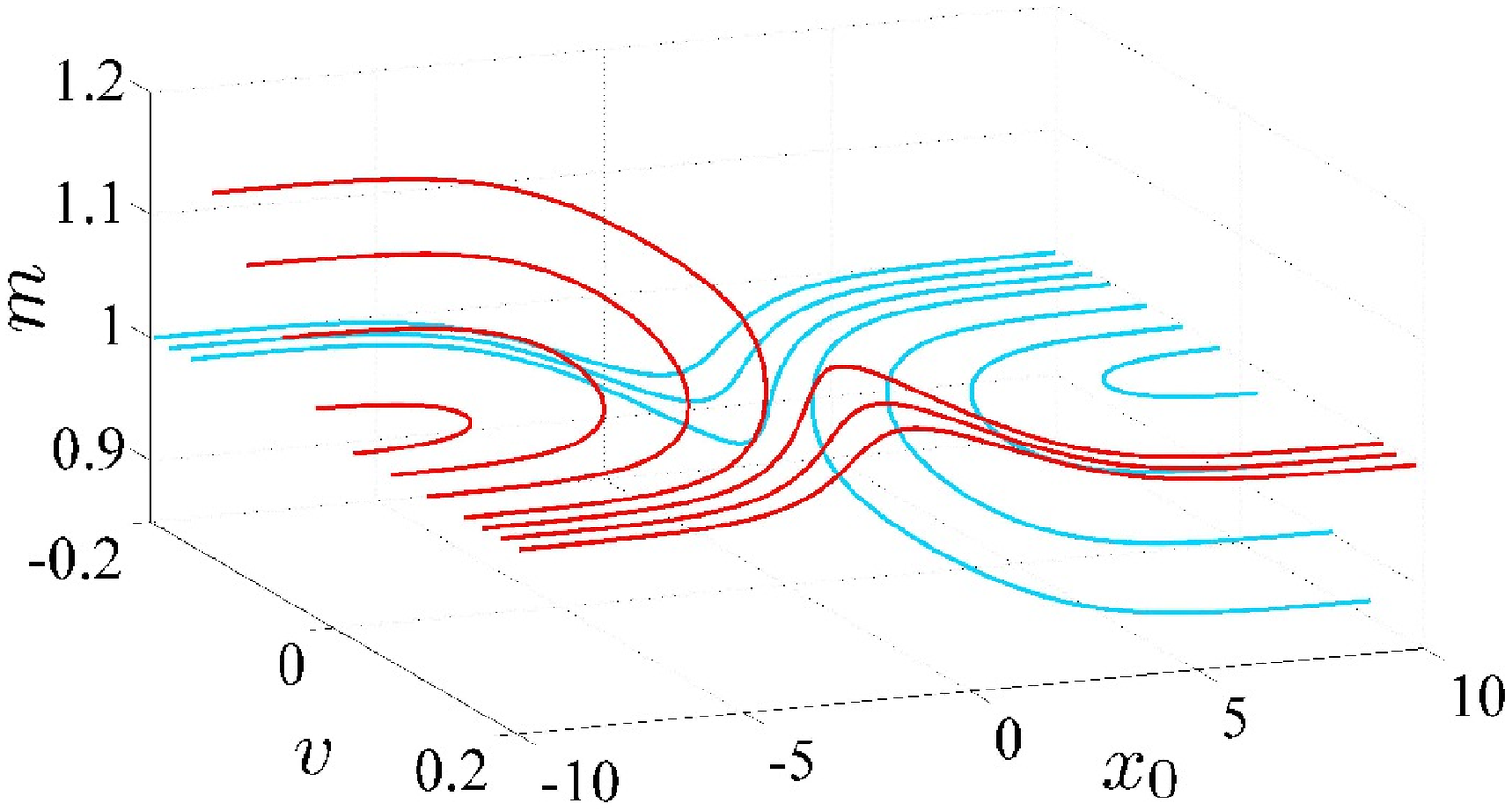}}}
 	\subfigure[]{\scalebox{\scl}{\includegraphics{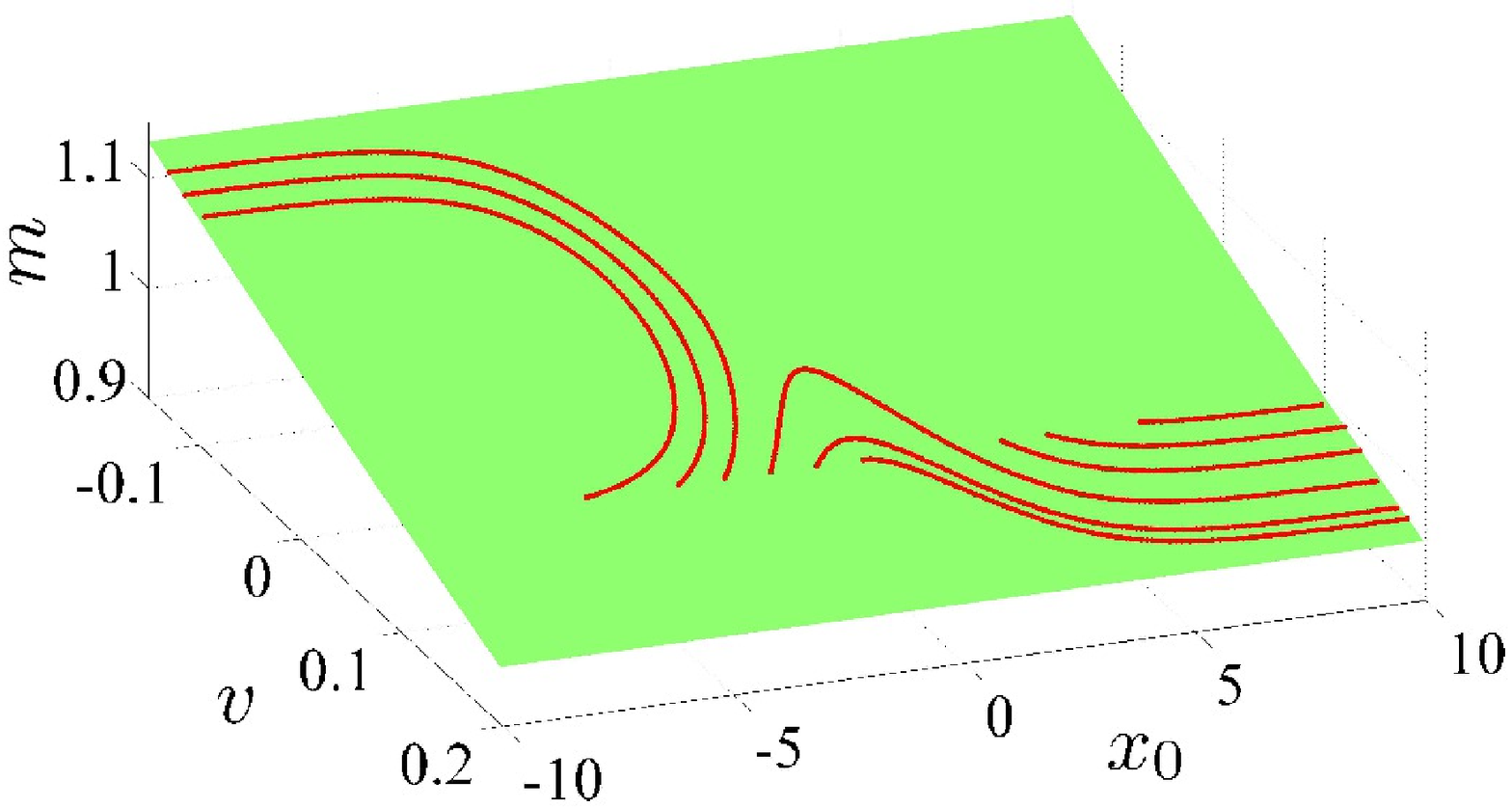}}}
 	\subfigure[]{\scalebox{\scl}{\includegraphics{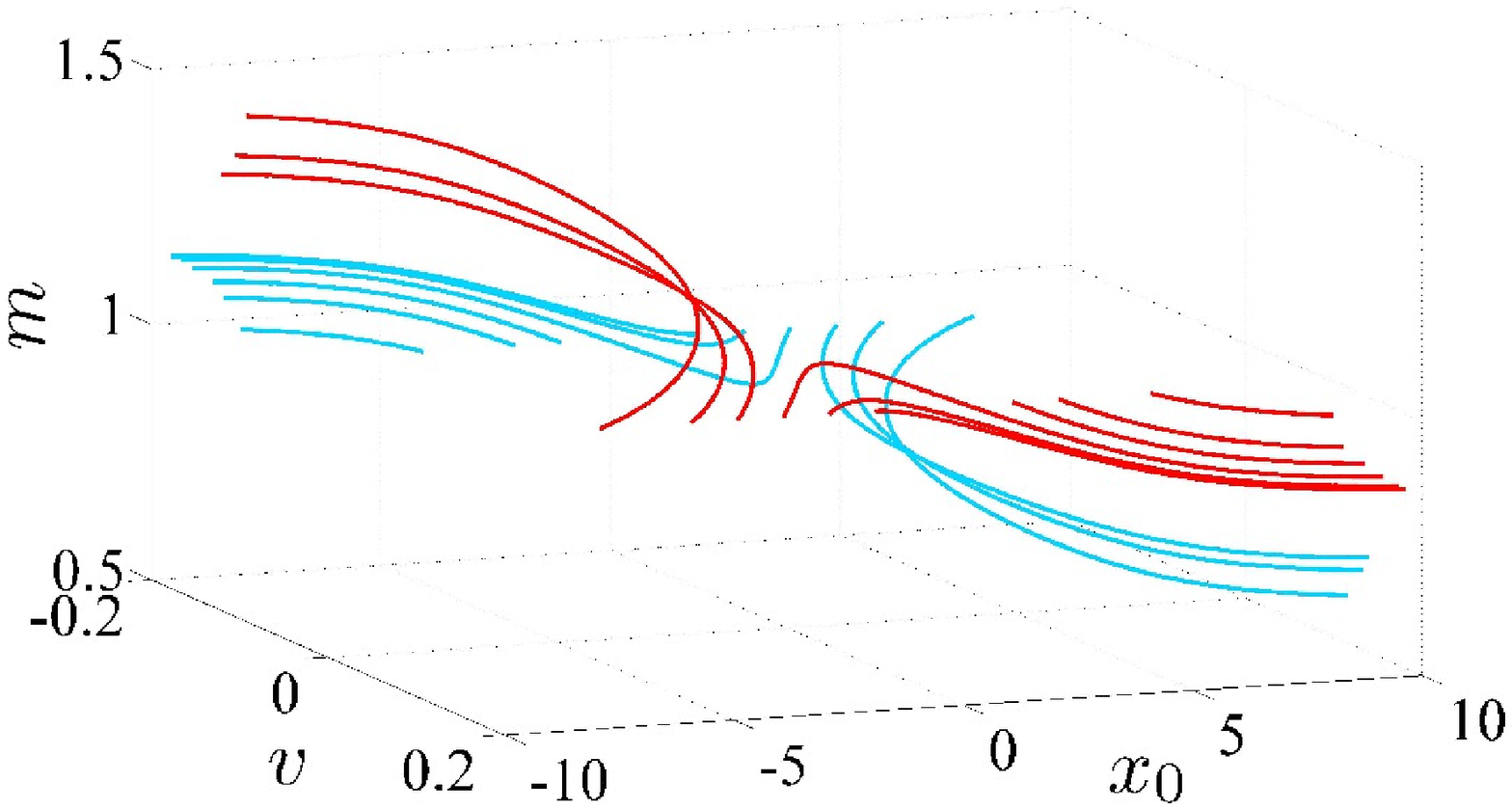}}}
 	\subfigure[]{\scalebox{\scl}{\includegraphics{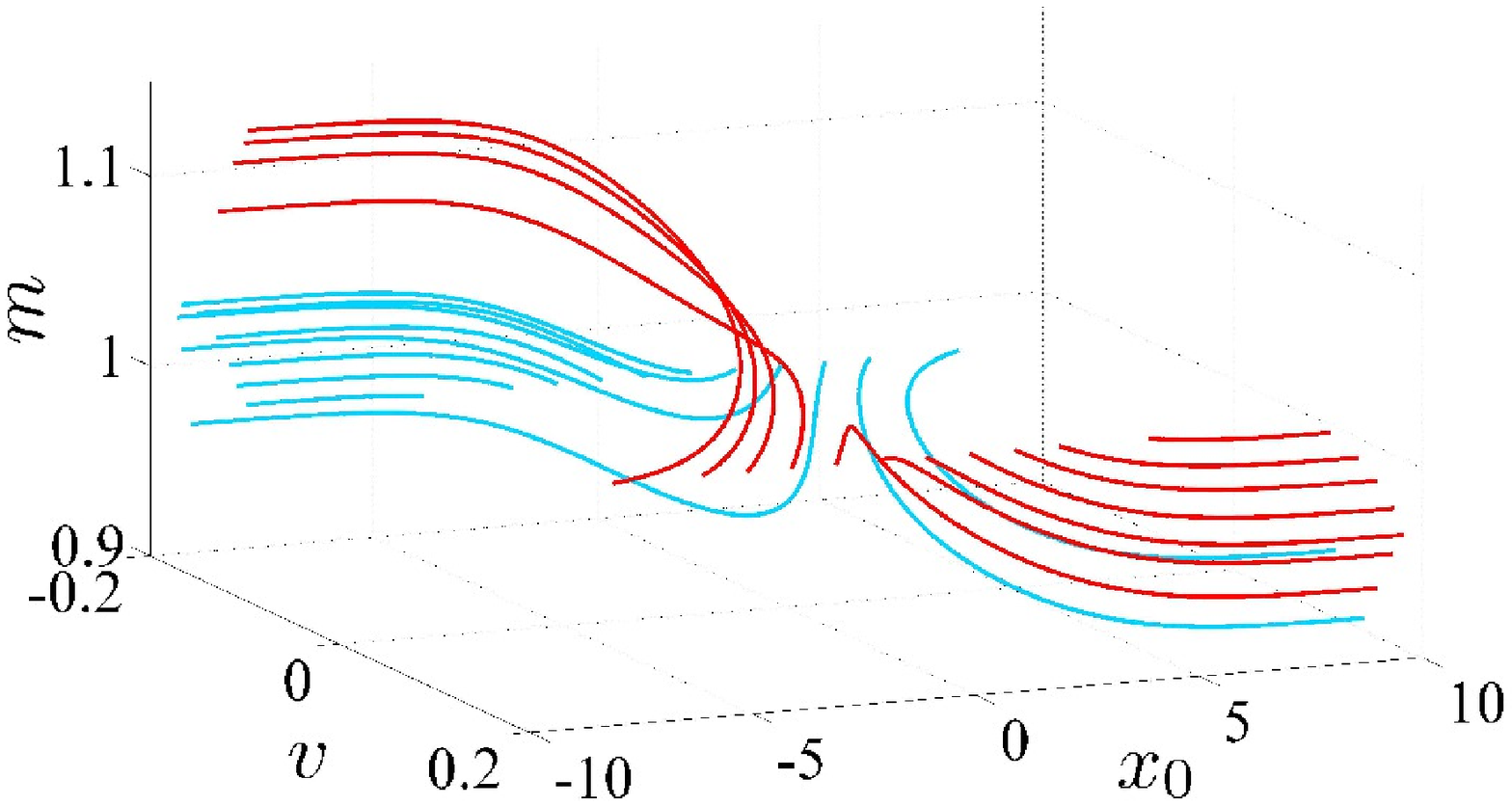}}}
 	\caption{(Color online) Phase space orbits of the effective particle model for a soliton with initial mass $m(0)=1$ in the localized potential (\ref{localized}) with $V_0=-0.01$, $W_0=|V_0|/2$ and $K_0=1$. (a) Potential: $L_0=1$, $\Delta x=0$, Initial conditions: $x_0(0)=-10$, $v(0)>0$ (red / dark gray) and $x_0(0)=10$, $v(0)<0$, (cyan / light gray); (b) Potential: $L_0=1$, $\Delta x=0$, Initial conditions: $v(0)=0.1$ (the two-dimensional surface (\ref{K}) is also shown); (c) Potential: $L_0=1/3$, $\Delta x=0$, Initial conditions: $x_0(0)<0$, $v(0)=0.1$ (red / dark gray) and $x_0(0)>0$, $v(0)=-0.1$, (cyan / light gray); (d) Potential: $L_0=1$, $\Delta x=-1$, Initial conditions: $x_0(0)<0$, $v(0)=0.1$ (red / dark gray) and $x_0(0)>0$, $v(0)=-0.1$, (cyan / light gray).}
	\label{Fig:3}
\end{center}
\end{figure} 

\begin{figure}[h]	
\begin{center}
	\subfigure[]{\scalebox{\scl}{\includegraphics{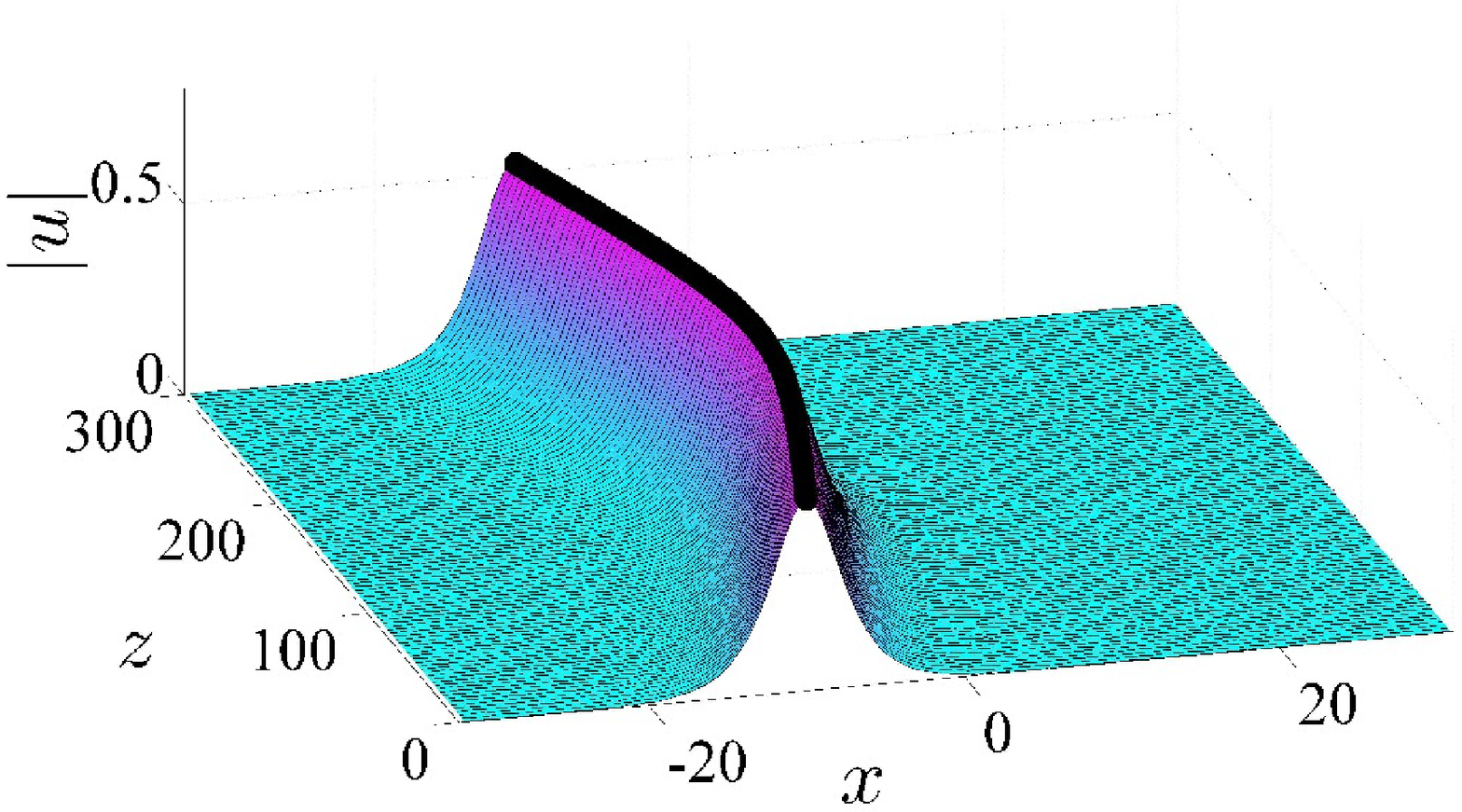}}}
 	\subfigure[]{\scalebox{\scl}{\includegraphics{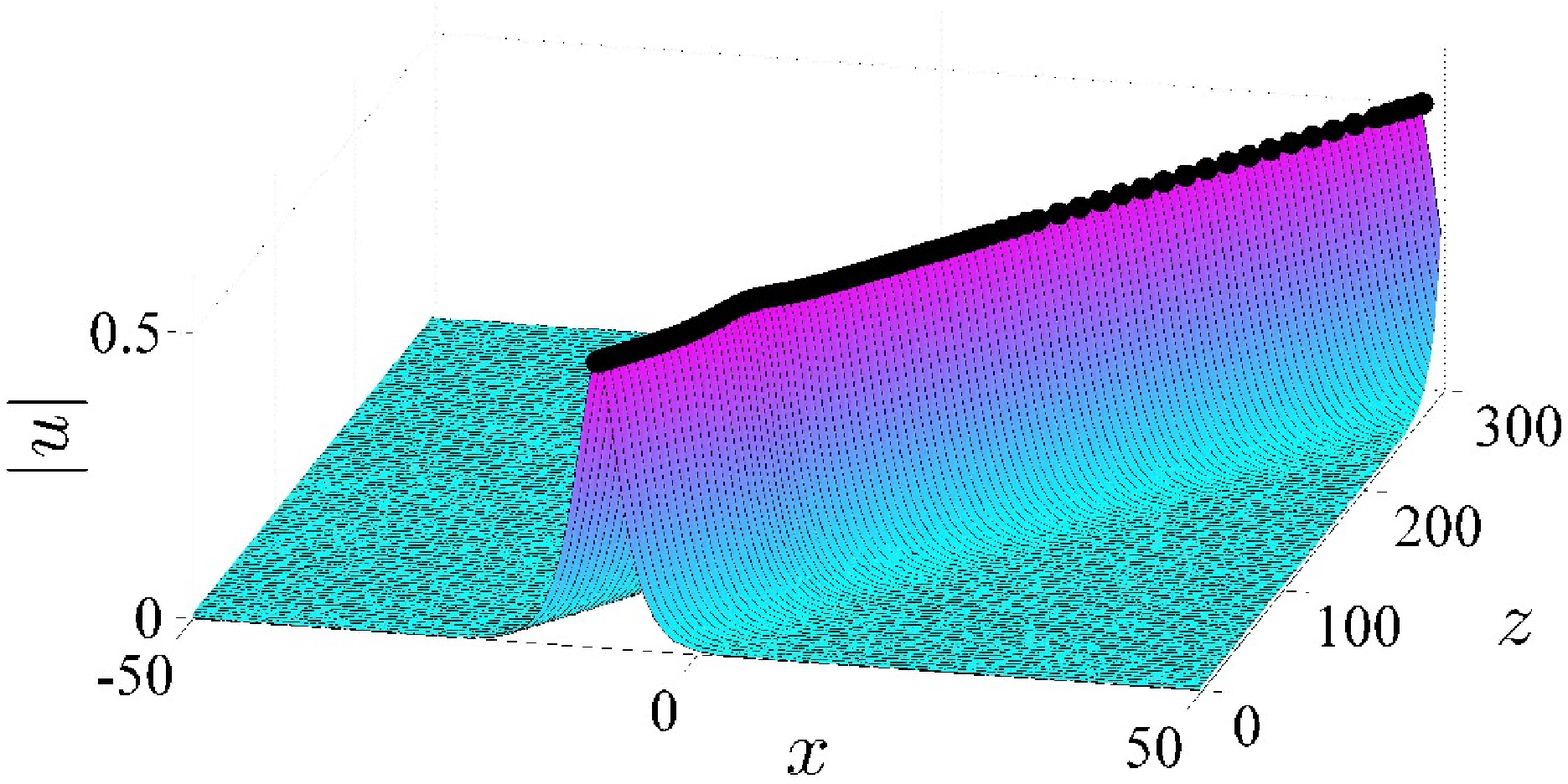}}}
 	\subfigure[]{\scalebox{\scl}{\includegraphics{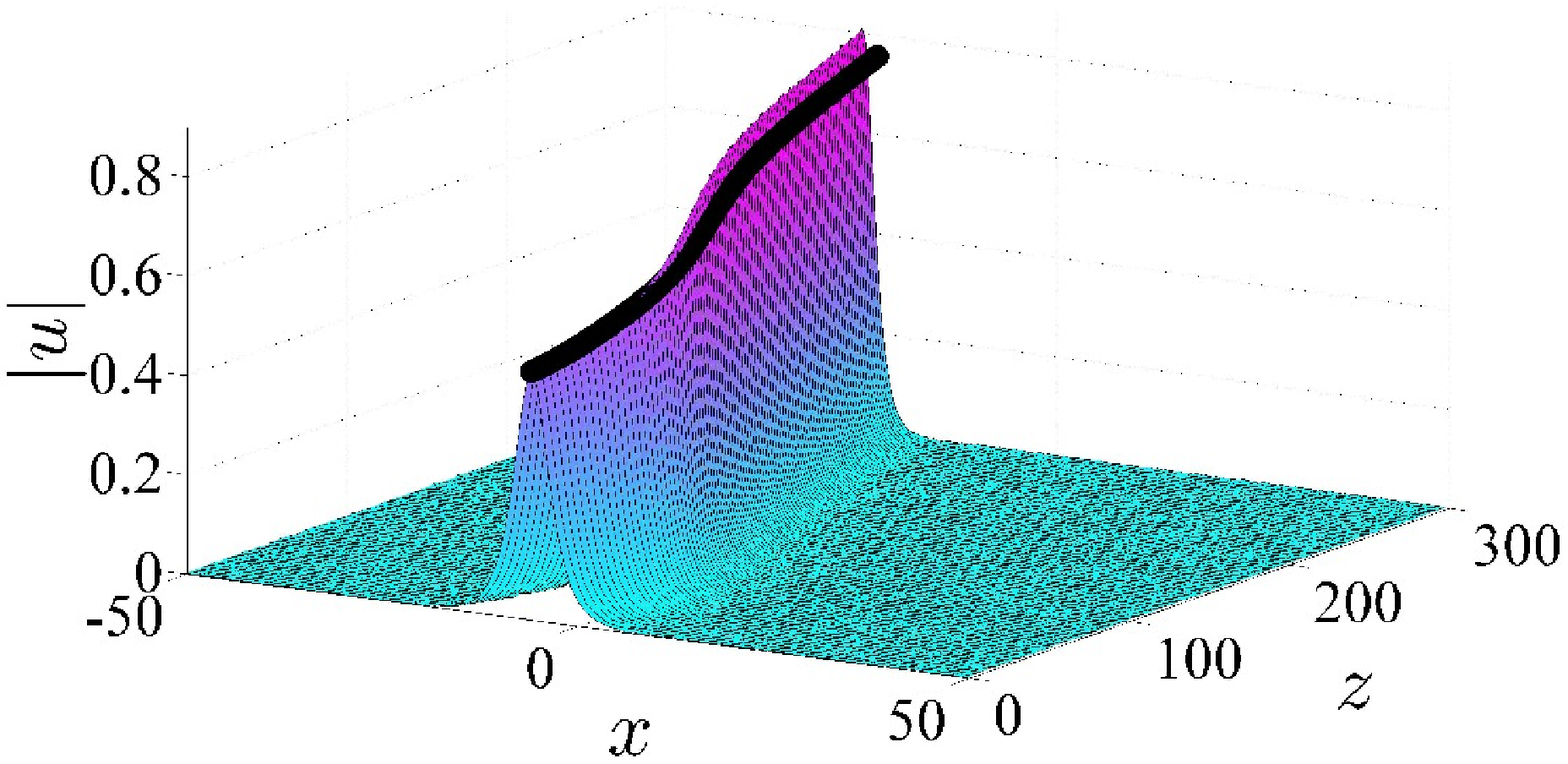}}}
 	\subfigure[]{\scalebox{\scl}{\includegraphics{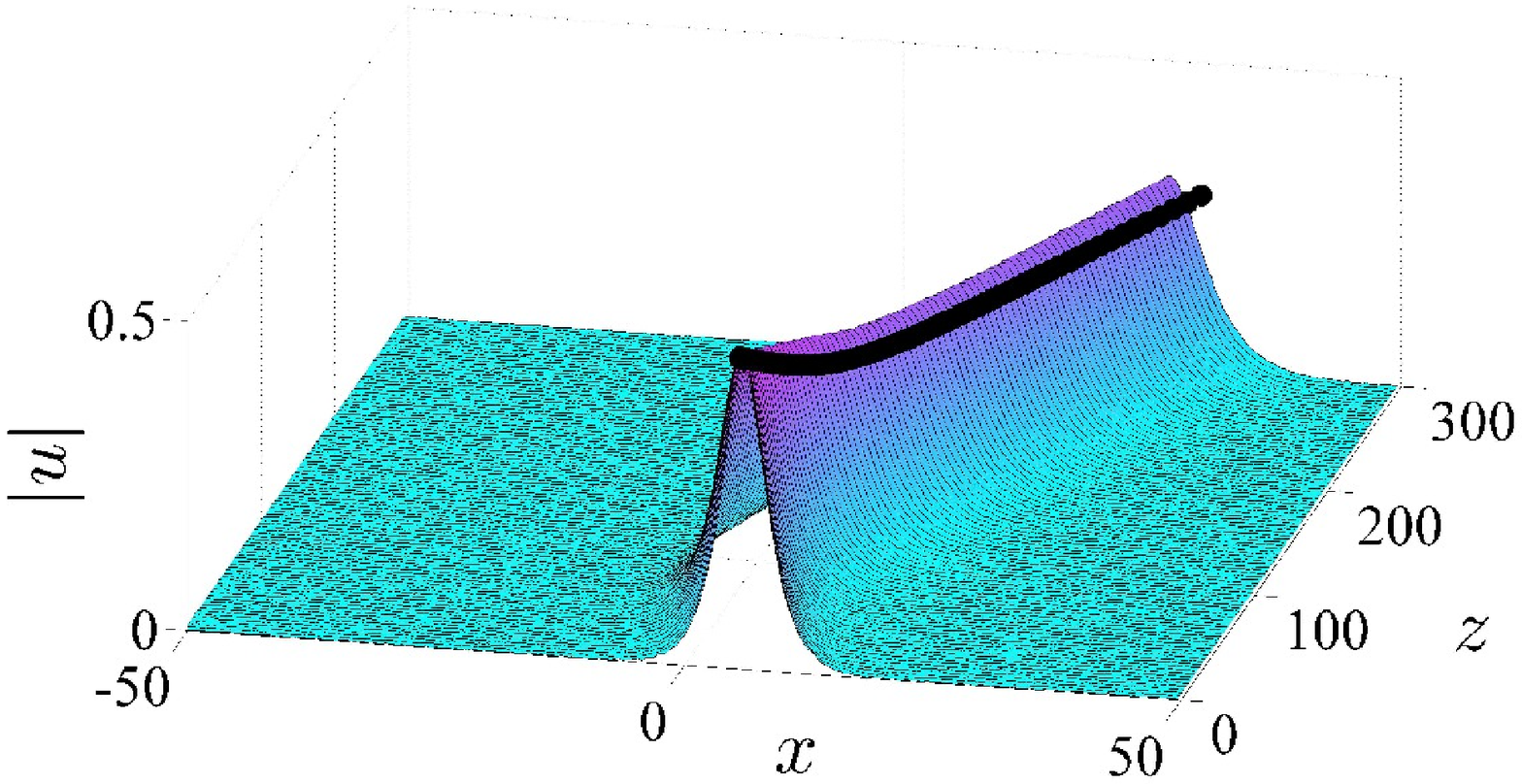}}}
 	\caption{(Color online) Soliton evolution under propagation as obtained from numerical simulations of the NLS equation (\ref{NLS}) and the effective particle model. The thick black line depicts the effective particle orbit $(x_0, v, m/2)$. (a) Potential: as in Fig. 3(a),(b), Initial conditions: $m(0)=1$, $x_0(0)=-10$, $v(0)=0.05$;  (b) Potential: as in Fig. 3(a),(b), Initial conditions: $m(0)=1$, $x_0(0)=-10$, $v(0)=0.2$; (c) Potential: as in Fig. 3(c), Initial conditions: $m(0)=1$, $x_0(0)=-6$, $v(0)=0.1$; (d) Potential: as in Fig. 3(c), Initial conditions: $m(0)=1$, $x_0(0)=6$, $v(0)=-0.1$.}
	\label{Fig:4}
\end{center}
\end{figure}

\section{Semi-infinite periodic potential}
A semi-infinite periodic potential corresponds to cases where a periodic structure is interfaced with a homogeneous part. In such case no spatial symmetry of the complex potential exists and the real and imaginary parts cannot be neither even nor odd functions. However, the condition (\ref{int_condition}) can still be fulfilled. We consider such a complex potential of the form 
\begin{eqnarray}
V(x)&=&V_0\left[1+\mbox{tanh}(a x)\right]\cos\left(K_0x\right) \nonumber \\
W(x)&=&W_0\left[1+\mbox{tanh}(a x)\right]\left\{\left[1-\mbox{tanh}(a x)\right]a\cos(L_0 x)-L_0\sin(L_0 x)\right\} \label{interface}
\end{eqnarray}
where the parameter $a$ determines the smoothness of the interface and $\partial V / \partial x=C W(x)$ with $C=V_0/W_0$ when $K_0=L_0$. Similarly to the case of a defect, the effective potential cannot be calculated analytically. Its form consist of a part of zero value extending to $-\infty$ and a part extending to $+\infty$ where it is periodic with the same characteristics as in the periodic case. In the vicinity of the interface it has a transitory form depending on the relation of the soliton mass with the additional spatial scale related to the smoothness of the interface $(a)$. \

The form of the effective potential for a soliton of mass $m=0.5$ is depicted in Fig. 5(a). It is remarkable that, due to the interface, a barrier is formed so that reflection can take place, not only for solitons traveling to the interface from the homogeneous side of the structure, but also for solitons coming from the periodic part of the structure for appropriate initial velocities. The formation of such potential barriers as well as potential wells in the homogeneous side close to the interface, depends strongly on the soliton mass, so that solitons of different mass can have qualitatively different evolution scenarios in the same structure. Phase space for the case where $K_0=L_0$ are depicted in Figs. 5(b) and (c) where the restriction to a two-dimensional surface for orbits with the same initial mass and velocity and various initial positions, due to the condition (\ref{int_condition}), is also shown. Characteristic cases of reflection of a soliton coming from the homogeneous side and a soliton trapped in an effective potential well formed in the homogeneous side are shown in Figs. 6(a) and (b). The interesting case of reflection of a soliton coming from the periodic side as well as soliton trapping at the periodic part, are shown in Figs. 6(c) and (d). Finally, for the case where $K_0 \neq L_0$, the condition (\ref{int_condition}) is not fulfilled and the phase space orbits are not restricted in a two-dimensional surface as shown in Fig. 5(d). In such cases we can have soliton reflection, transmission as well as soliton trapping with a continuously increasing or decreasing mass depending on the local value of the gain/loss, similarly to the cases depicted in Figs. 1(d) and (f).  

\begin{figure}[h]	
\begin{center}
	\subfigure[]{\scalebox{\scl}{\includegraphics{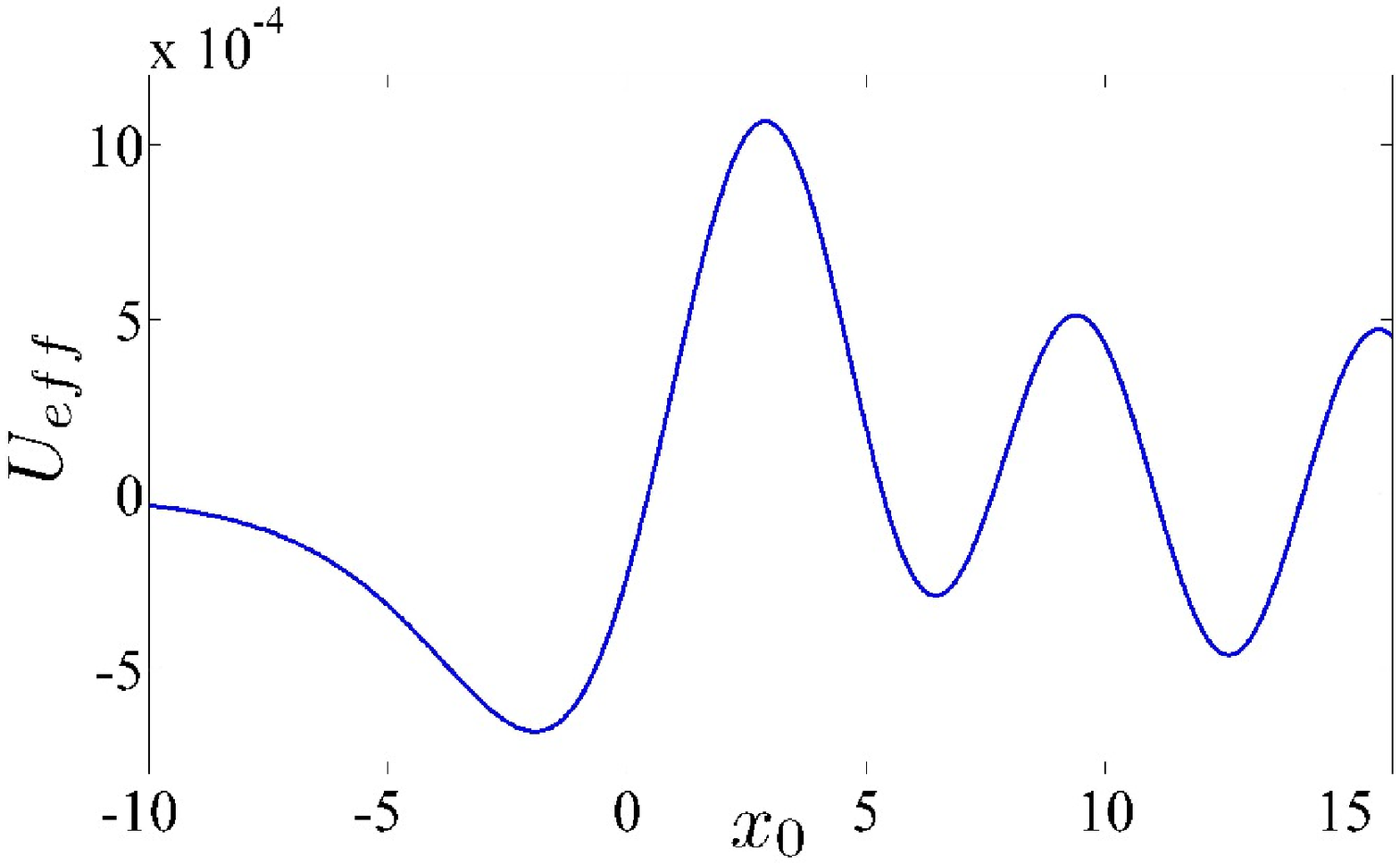}}}
 	\subfigure[]{\scalebox{\scl}{\includegraphics{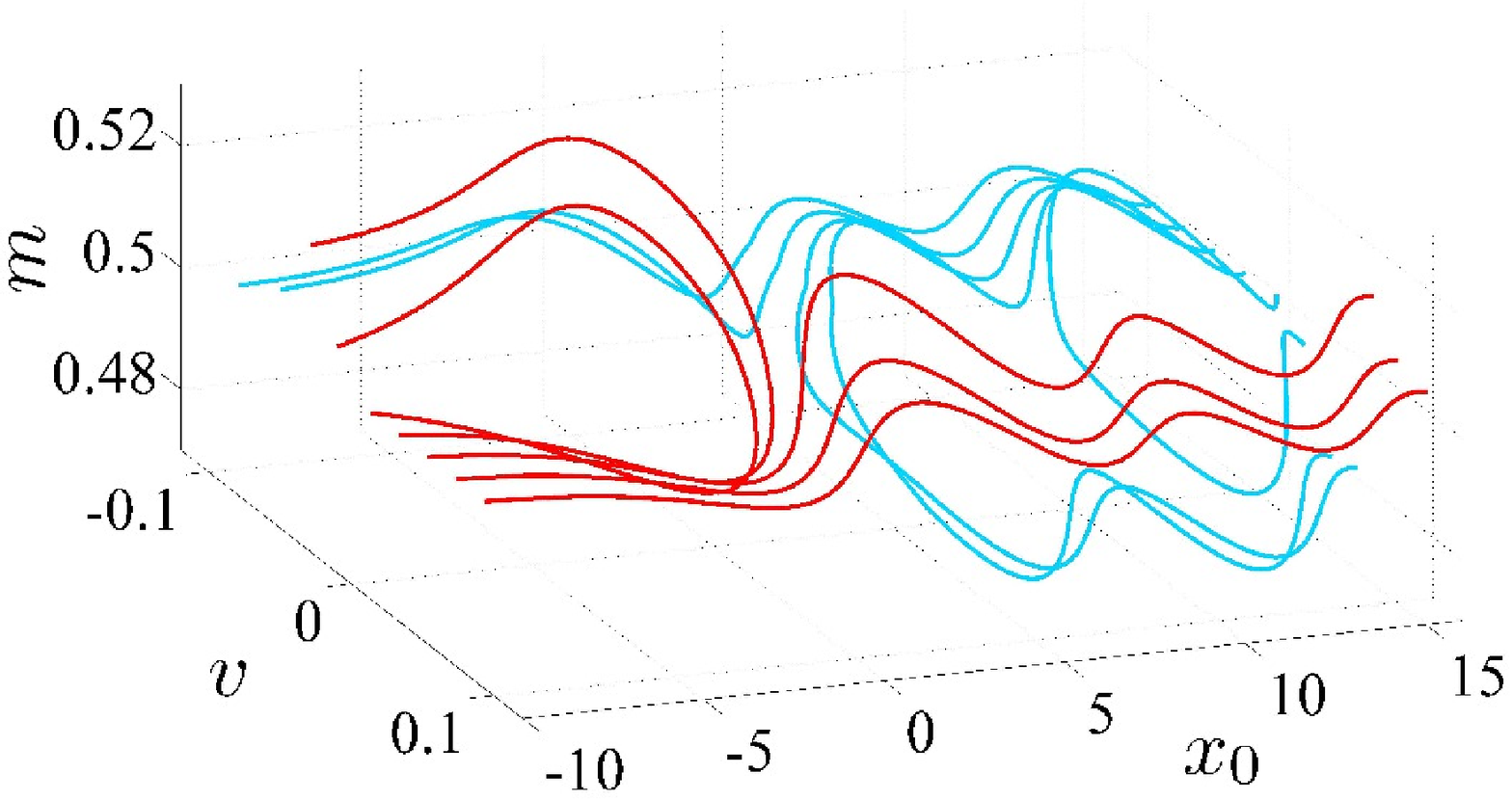}}}
 	\subfigure[]{\scalebox{\scl}{\includegraphics{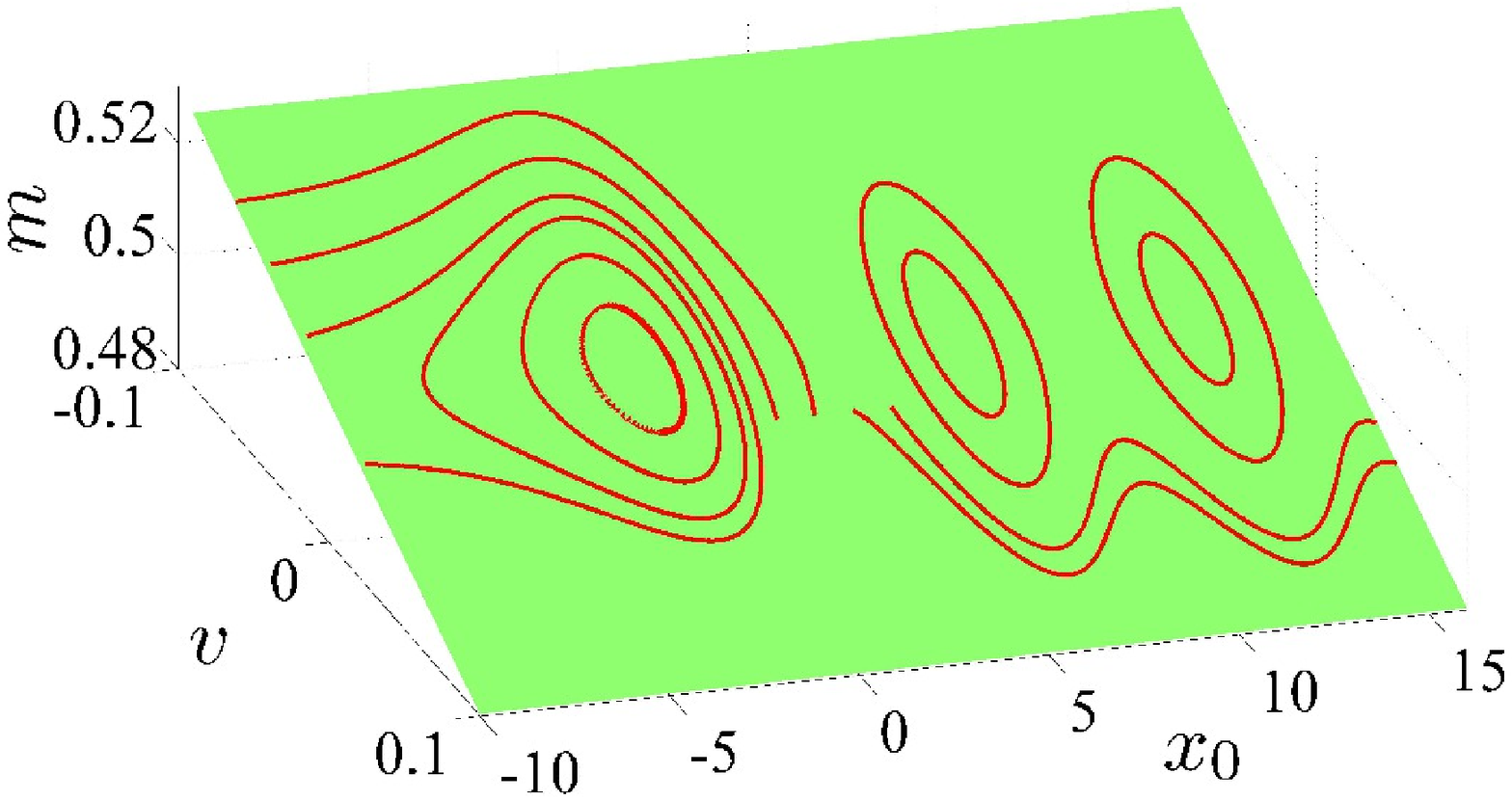}}}
 	\subfigure[]{\scalebox{\scl}{\includegraphics{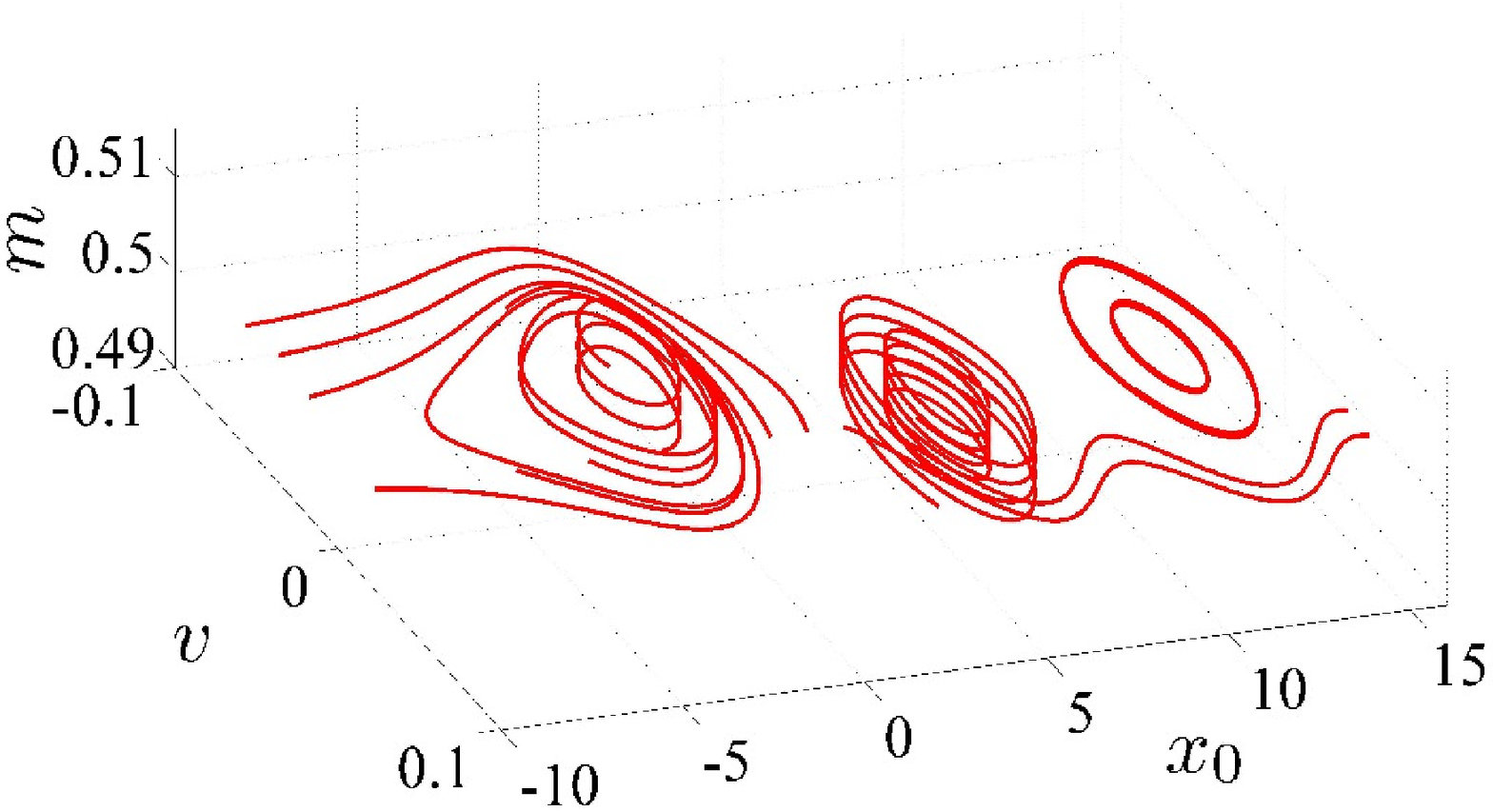}}}
 	\caption{(Color online) Effective potential and phase space orbits of the effective particle model for a soliton with initial mass $m(0)=0.5$ in the semi-infinite periodic potential (\ref{interface}) with $V_0=0.01$, $W_0=V_0/2$, $a=1$ and $K_0=1$. (a) Effective potential; (b) Potential: $L_0=1$, Initial conditions: $x_0(0)=-10$, $v(0)>0$ (red / dark gray) and $x_0(0)=16$, $v(0)<0$, (cyan / light gray); (c) Potential: $L_0=1$, Initial conditions: $v(0)=0.02$ (the two-dimensional surface (\ref{K}) is also shown); (d) Potential: $L_0=2$, Initial conditions: $v(0)=0.02$.}
	\label{Fig:5}
\end{center}
\end{figure} 

\begin{figure}[h]	
\begin{center}
	\subfigure[]{\scalebox{\scl}{\includegraphics{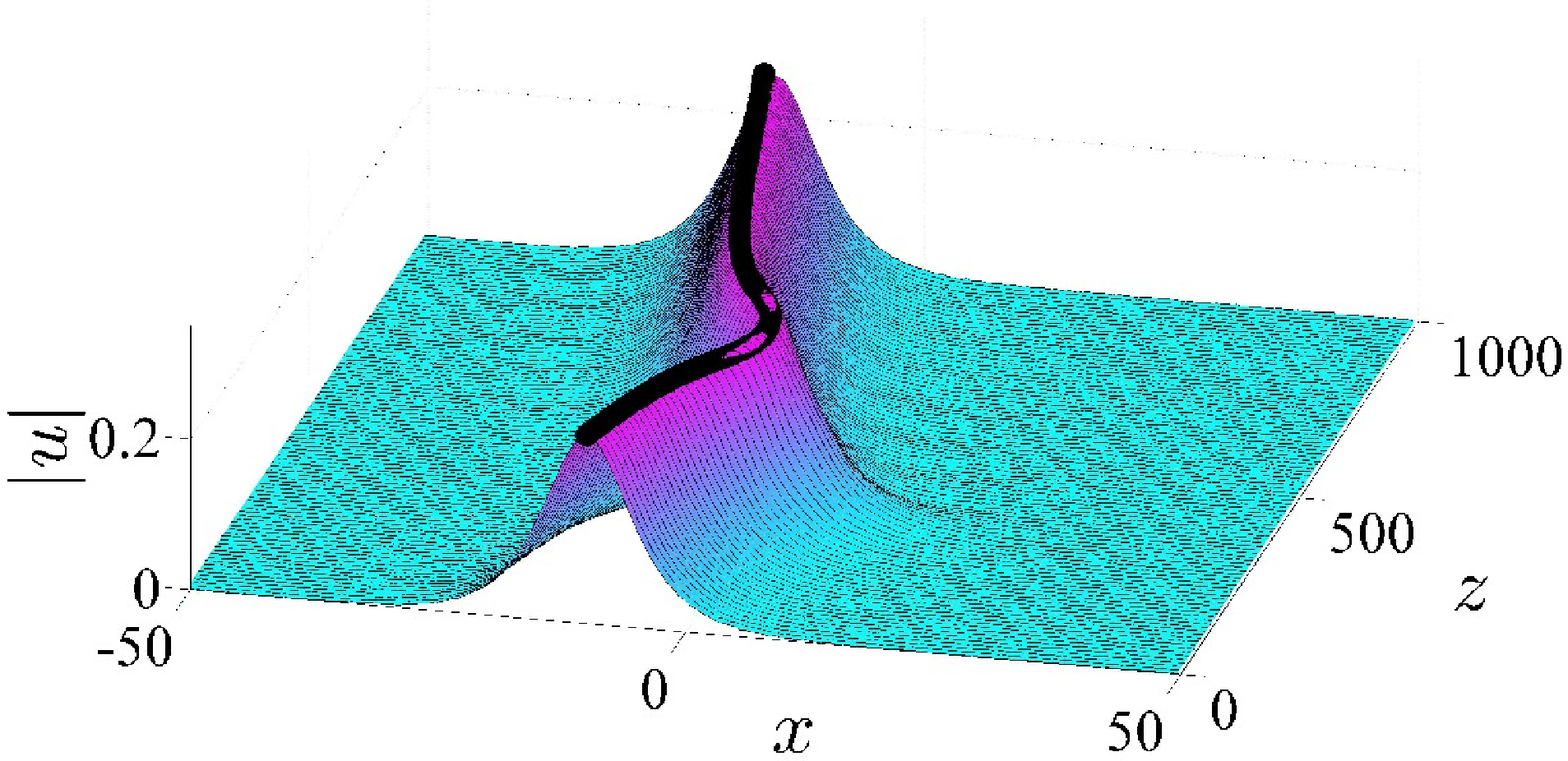}}}
 	\subfigure[]{\scalebox{\scl}{\includegraphics{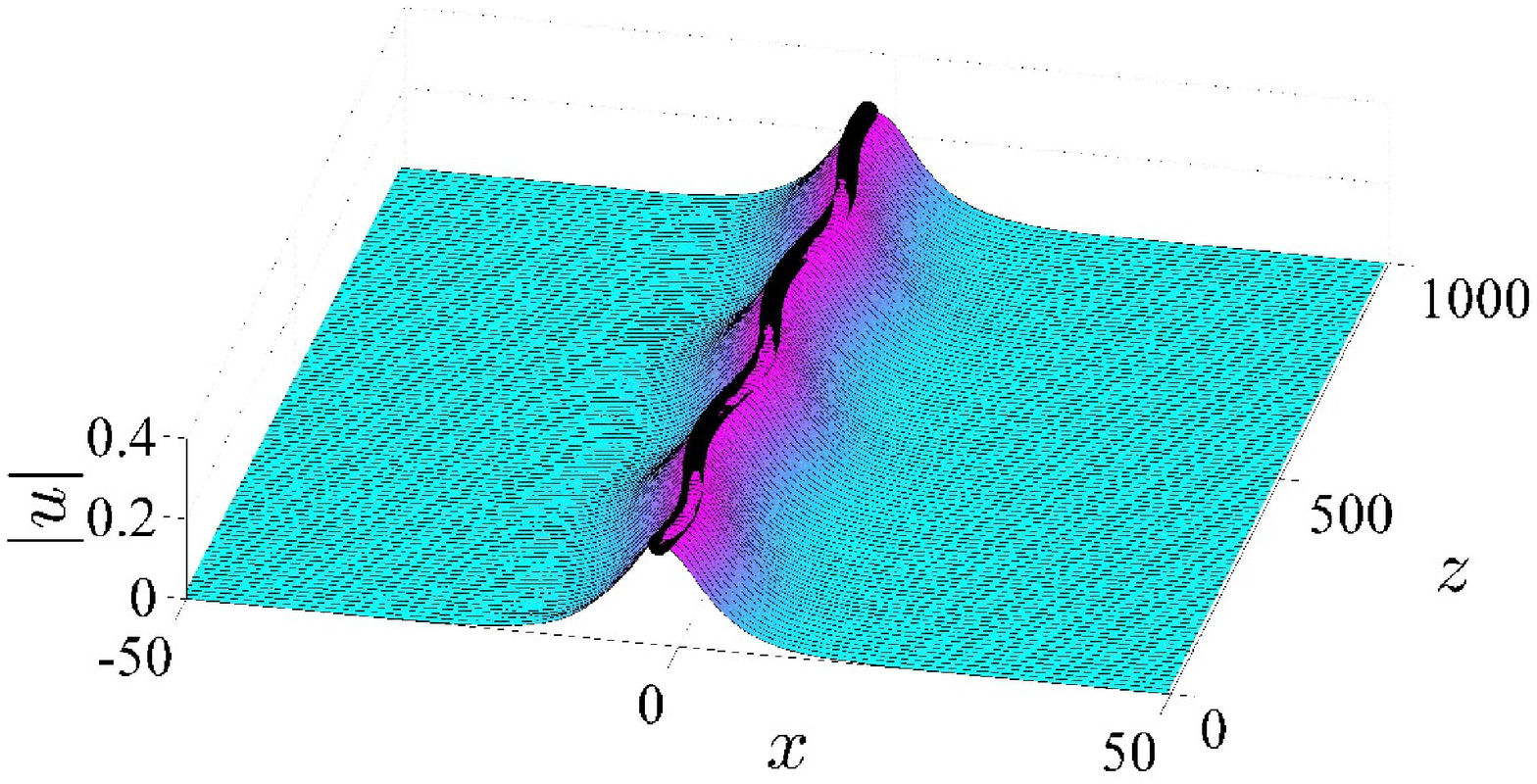}}}
 	\subfigure[]{\scalebox{\scl}{\includegraphics{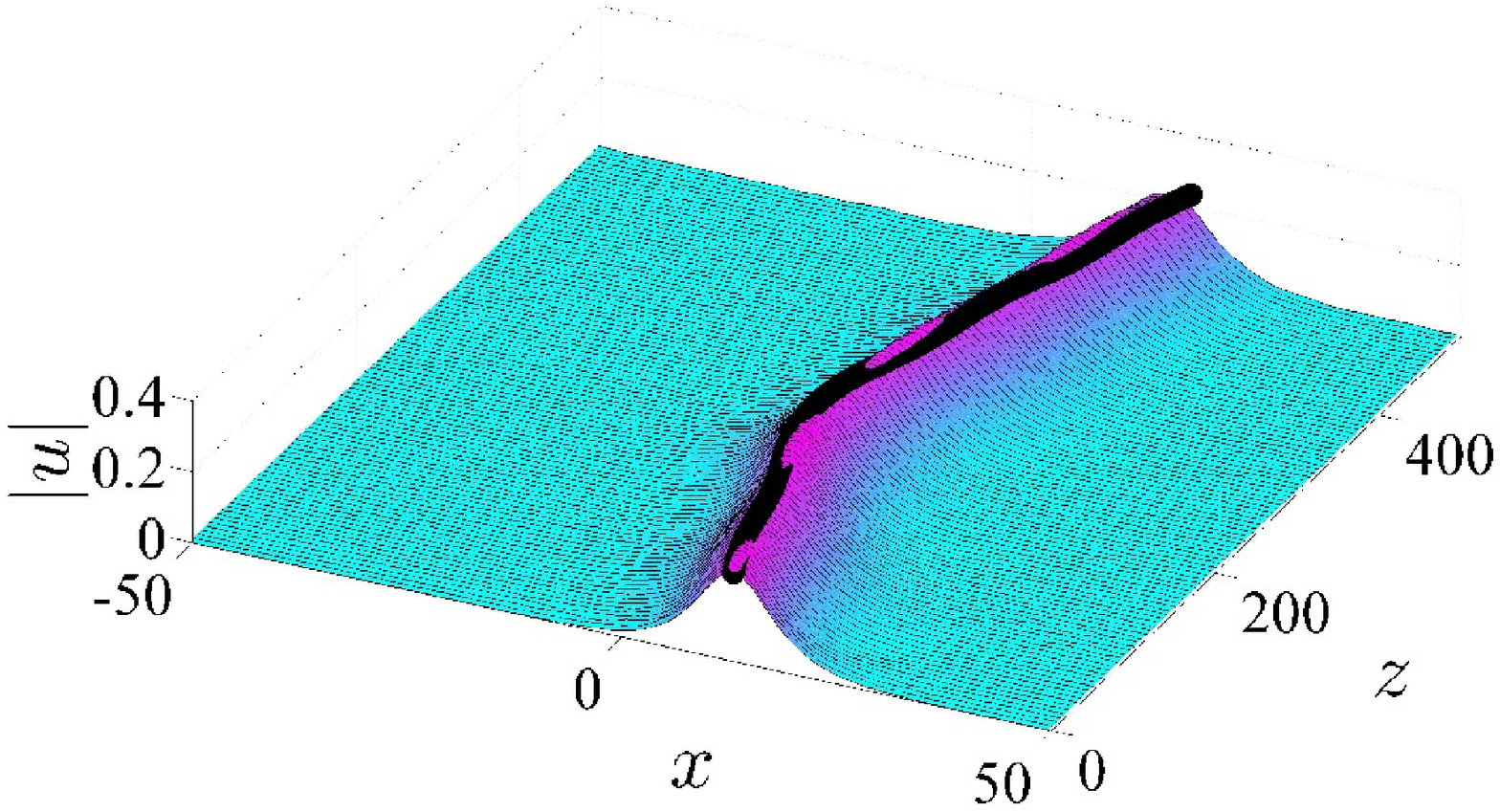}}}
 	\subfigure[]{\scalebox{\scl}{\includegraphics{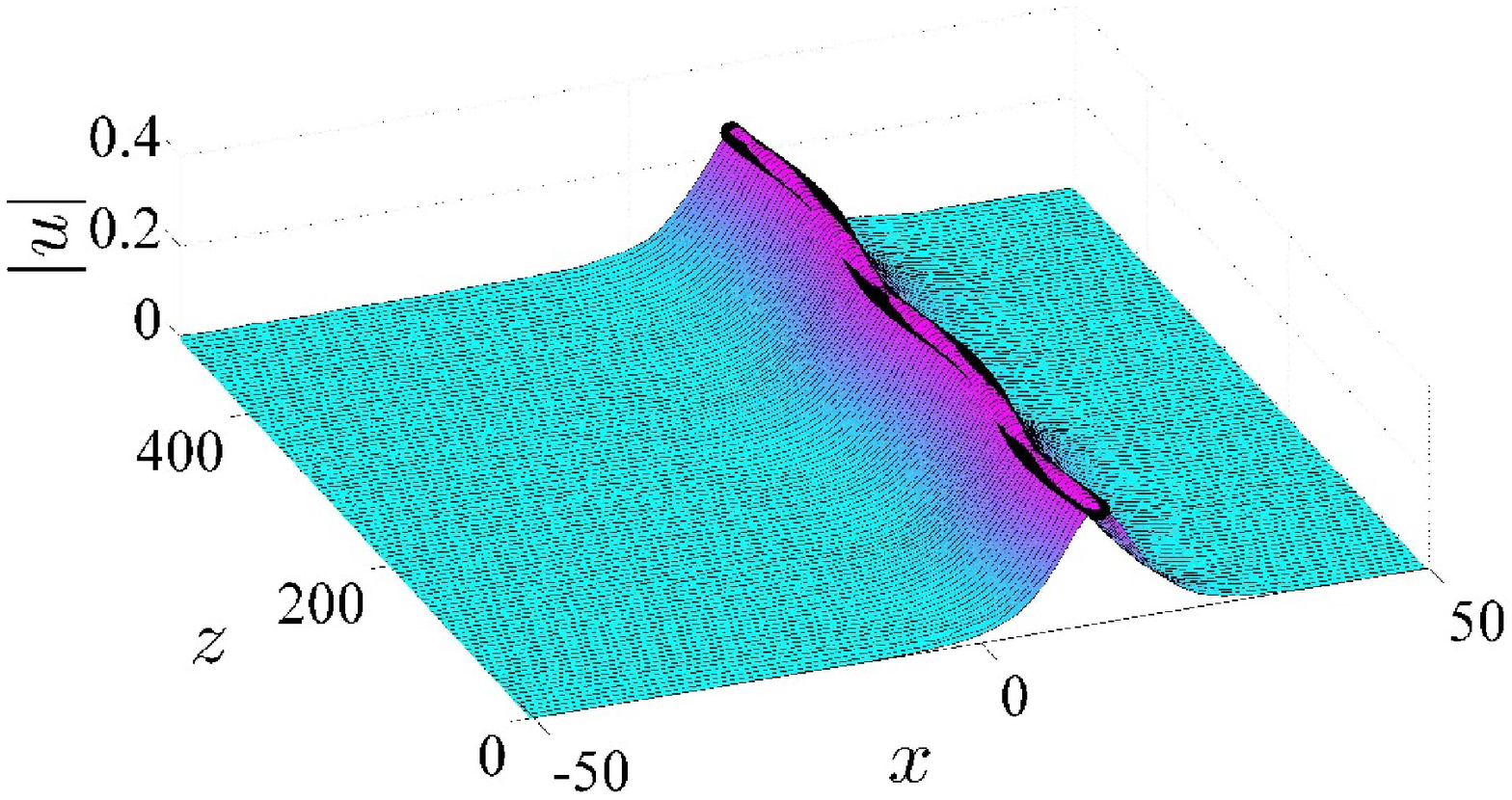}}}
 	\caption{(Color online) Soliton evolution under propagation as obtained from numerical simulations of the NLS equation (\ref{NLS}) and the effective particle model. The thick black line depicts the effective particle orbit $(x_0, v, m/2)$. Potential: as in Fig. 5(a)-(c), Initial conditions: $m(0)=0.5$ and (a) $x_0(0)=-10$, $v(0)=0.02$, (b) $x_0(0)=-2$, $v(0)=0.02$, (c) $x_0(0)=13$, $v(0)=-0.07$, (d) $x_0(0)=13$, $v(0)=-0.03$. }
	\label{Fig:6}
\end{center}
\end{figure}

\section{Summary and Conclusions}
Soliton propagation under the presence of refractive index and gain/loss inhomogeneity has been studied in terms of a NLS equation with a complex linear potential. An effective particle approach has been utilized in order to systematically investigate soliton dynamics in the three-dimensional phase space consisting of soliton center position, transverse velocity and mass. The existence of fixed points for stationary soliton propagation has been shown to correspond to the necessary conditions for $\mathcal{PT}$- symmetry, restricting the real and the imaginary part of the complex potential to the class of even and odd functions, respectively. An invariant of the soliton motion and mass variation has been derived under a condition mutually restricting the profiles of the real and the imaginary part of the complex potential. \

The study has been focused on relatively weak potentials for which solitons are quite mobile, in contrast to cases of strong potentials where solitons are deeply trapped. Therefore, the profiles of the solitons remain simple whereas their dynamics are shown to have a very rich set of interesting features. The phase space analysis has been used in order to provide intuitive understanding and to categorize qualitatively different soliton evolution scenarios in a large variety of complex potentials. Infinite periodic, localized  and semi-infinite periodic potentials have been considered, with their profiles being either symmetric or non-symmetric. The additional degree of freedom, corresponding to mass variation, has been shown to result in space and velocity symmetry breaking and non-reciprocity of soliton dynamics. Direct simulations of the original NLS equation have been shown in excellent agreement with the effective particle model. Moreover, they have confirmed the nondestructive character of the zero background instability for relatively weak potentials in finite propagation distances, that are of interesting to realistic applications. \ 

The approach and the results presented in this work can be directly extended to cases of a nonlinear complex potential, combinations of linear and nonlinear complex potentials and any type of spatial potential profile including lattices, superlattices, interfaces and localized defects, as well as to two-dimensional potentials. The richness of the dynamical features of soliton propagation along with the freedom of utilizing a large variety of complex potentials is very promising for soliton control in appropriately engineered structures.

\section*{Acknowledgments}
The author acknowledges support by the research project NWDCCPS implemented within the framework of the Action "Supporting Postdoctoral Researchers" of the Operational Program "Education and Lifelong Learning" (Action's Beneficiary: General Secretariat for Research and Technology), and is co-financed by the European Social Fund (ESF) and the Greek State.

\end{document}